\documentclass[a4paper,11pt]{article}
\pdfoutput=1 

\usepackage{jcappub} 

\usepackage[T1]{fontenc} 

\usepackage{graphicx}
\usepackage{amssymb}
\usepackage{epstopdf}
\usepackage{morefloats}
\usepackage{parskip}
\usepackage[section]{placeins}
\usepackage{float}
\usepackage{afterpage}
\usepackage{leftidx}
\usepackage[section]{placeins}
\usepackage{tikz}
\usepgflibrary{arrows}
\usepackage{mathtools}
\usepackage{caption}
\usepackage{subcaption}
\usepackage{array, multirow}
\usepackage{color}
\usepackage{url}
\usepackage{hyperref}
\usepackage{textcomp}
\usepackage{mathrsfs}
\usepackage{amssymb}
\usepackage{tablefootnote} 
\usepackage{booktabs}
\usepackage{multirow}
\newcommand{\be}{\begin{equation}}
\newcommand{\ee}{\end{equation}}
\newcommand{\beq}{\begin{eqnarray}}
\newcommand{\eeq}{\end{eqnarray}}

\def\nue{\mathrel{{\nu_e}}}
\def\numu{\mathrel{{\nu_\mu}}}
\def\nutau{\mathrel{{\nu_\tau}}}
\def\nux{\mathrel{{\nu_x}}}

\def\barnue{\mathrel{{\bar \nu}_e}}
\def\barnumu{\mathrel{{\bar \nu}_\mu}}
\def\barnutau{\mathrel{{\bar \nu}_\tau}}

\def \lta {\mathrel{\vcenter{\hbox{$<$}\nointerlineskip\hbox{$\sim$}}}}
\def \gta {\mathrel{\vcenter{\hbox{$>$}\nointerlineskip\hbox{$\sim$}}}}

\def\t13{\mathrel{{\theta_{13}}}}
\def\y12{\mathrel{{\tan^2 \theta_{12}}}}
\def\c2{\mathrel{{\chi^2 }}}

\def \msun	 	{M_{\odot}}

\newcommand{\n}{neutrino}
\newcommand{\ns}{neutrinos}
\newcommand{\sn}{supernova}
\newcommand{\sne}{supernovae}
\newcommand{\skg}{SuperK-Gd}
\newcommand{\sk}{Super-Kamiokande}
\newcommand{\ck}{Cherenkov}
\newcommand{\jn}{JUNO}
\newcommand{\df}{DSNB}
  \newcommand{\nsf}{NSFC}
   \newcommand{\dbh}{BHFC}
    \newcommand{\bh}{BH}
     
     \newcommand{\snr}{SNR}
    
\title{\boldmath Diffuse neutrinos from luminous and dark supernovae: prospects for upcoming detectors at the $\mathcal{O}$(10) kt scale}

\author[a,1]{Alankrita Priya} \note{Corresponding author. This is an author-created, un-copyedited version of an article published in the Journal of Cosmology and Astroparticle Physics. IOP Publishing Ltd is not responsible for any errors or omissions in this version of the manuscript or any version derived from it. The Version of Record is available online at https://doi.org/10.1088/1475-7516/2017/11/031.}

\author[a]{and Cecilia Lunardini}

\affiliation[a]{Department of Physics, Arizona State University, Tempe, Arizona 85287, USA}

\emailAdd{Alankrita.Priya@asu.edu}
\emailAdd{Cecilia.Lunardini@asu.edu}
\abstract{We estimate the Diffuse Supernova Neutrino Background (DSNB) using the simulation results for neutron star-forming and  black hole-forming stellar collapses
  from the Garching group. Scenarios with different distributions of black-hole forming collapses with the progenitor mass are discussed, and the uncertainty on the cosmological rate of collapses is included.  The $\barnue$ component of the DSNB above 11 MeV of energy is found to be $\phi \simeq (1.4 - 3.7)~{\rm cm^{-2} s^{-1}}$; the contribution of black hole-forming collapses could dominate the flux above $\sim 25$ MeV.   
 We calculate the  potential of detecting the DSNB at SuperK-Gd and JUNO, in about a decade-long period of operation. We find that, in our model, it is likely that a significant excess above the background will be obtained at JUNO, while detection will be more difficult at SuperK-Gd.
  The potential when the two experimental results are examined jointly is discussed as well.
  We also consider an example of 
   a future ${\mathcal O}(10)$ kt 
slow liquid scintillator detector, and show that there the odds of detection are very good. Our results motivate experimental efforts in reducing the backgrounds due to neutral current scattering of atmospheric neutrinos in SuperK-Gd. 
}

\begin{document}
\maketitle
\flushbottom

\section{Introduction} 
\label{sec:intro}

Neutrinos from core collapse supernovae are unique messengers of late stellar evolution and of nuclear and particle physics at the extreme conditions near the collapsing core.  A medium- or high-statistics observation of these neutrinos could answer many fundamental questions ranging from the equation of state of nuclear matter to the existence of new particles and interactions. 

After the low statistics detection from SN1987A \cite{Hirata:1987aa,PhysRevLett.58.1494, Alekseev:1987ej}, the next opportunity to detect \sn\ \ns\ could be offered by
the Diffuse Supernova Neutrino Background (\df), the diffuse flux from all the \sne\ in the universe \cite{NYAS:NYAS319,Krauss:1984aa}. This flux is constant in time, and therefore progress towards its observation is essentially technologically driven.  Current upper limits on the \df\ \cite{PhysRevD.85.052007} are close to
theoretical predictions (see, e.g., \cite{TOTANI1995367,MALANEY1997125,Hartmann1997137,PhysRevD.62.043001,ANDO2003307,1475-7516-2005-04-017,Lunardini2006190}), leading to the hope that a detection might be achieved at the next generation of \n\ observatories.  

The \df\ is especially interesting because it probes the entire population of collapsing stars, in its diversity and cosmological evolution. A striking illustration of this is the idea that the \df\ might carry the imprint of black hole formation \cite{Lunardini:2009aa,PhysRevD.85.043011,doi:10.1146/annurev.nucl.010909.083331,PhysRevD.79.083013,Lien:2010aa}
in the form of a hotter component due to {\it failed \sne}, stars that collapse directly into black holes (BH) without an explosion.  The possibility to learn about the birth of black holes using \ns\ opens interesting interdisciplinary connections with studies of  General Relativity and with the new frontier of gravitational wave detection from black hole mergers \cite{PhysRevLett.116.061102}. 

Studies of the contribution from failed \sne\ (or  Black Hole-Forming Collapses, \dbh) on the \df\ so far  \cite{doi:10.1146/annurev.nucl.010909.083331, Lunardini:2009aa, 0004-637X-790-2-115, Nakazato:2008aa, Lien:2010aa,0004-637X-804-1-75} have captured the basic elements, namely the hotter energy spectrum (compared to Neutron Star-Forming Collapses, \nsf) and a $ {\mathcal O}(10\%)$ fraction of collapses that directly produce a black hole.  The energy spectra of the individual \n\ flavors were either parameterized phenomenologically, or taken from pioneering numerical simulations of direct black hole formation for two different equations of state \cite{Sumiyoshi:2006aa, Nakazato:2008aa,doi:10.1063/1.1419594,0004-637X-804-1-75}. The diffuse flux from \dbh\ was estimated considering a single progenitor as representative of the entire population of failed \sne, and the fraction of \dbh\ was modeled either as a constant, or as a redshift-dependent parameter \cite{0004-637X-790-2-115,0004-637X-804-1-75}, reflecting the metallicity evolution of the stellar population \cite{0004-637X-804-1-75}.
Detectability studies (e.g., \cite{0004-637X-790-2-115,PhysRevD.85.043011}) 
mostly 
considered a vision where current experiments would be succeeded  by a  new generation of detectors at  0.1-1 Mt mass,  where the \df\ could be detected above backgrounds with medium-high statistics.

In the recent years, the situation has matured considerably.   On the theory front, detailed studies have appeared on how the outcome of the collapse (black hole  or neutron star formation) depends on the stellar structure of the progenitor star \cite{0004-637X-762-2-126, 0004-637X-757-1-69, 0004-637X-801-2-90,0004-637X-818-2-124}.  The \n\ spectra have been modeled for a number of progenitors of varying masses 
and metallicity \cite{phdthesis, Nakazato:2008aa, Sumiyoshi:2006aa}, and incorporating 
convection and detailed state-of-the art microphysics \cite{phdthesis, Mirizzi:2015eza}. 
Astronomical observations have progressed, further supporting the failed \sn\ hypothesis. It has been discussed how \bh\  formation can naturally explain the problem of missing red supergiant stars \cite{doi:10.1093/mnrasl/slu146,0004-637X-785-1-28}. Recently, a failed \sn\ candidate has been identified, as a star that has disappeared from the sky \cite{0004-637X-684-2-1336}.

On the experimental front, a concrete path forward has emerged. While Mt-scale experiments remain a goal for the distant future, new, medium-scale detectors are currently being built.  The upcoming  Jiangmen Underground Neutrino Observatory (\jn) \cite{0954-3899-43-3-030401} will be the largest liquid scintillator detector ever realized (17 kt fiducial mass), with unprecedented energy resolution. Detailed, realistic models of the backgrounds of \df\ searches at \jn\ have been published recently  \cite{0954-3899-43-3-030401},  and have stimulated ideas on how to further improve the potential of liquid scintillator   for \df\ detection \cite{Wei2017255}. 
The even larger \skg\ is the approved Gadolinium-based upgrade of the \sk\ detector \cite{Beacom:2004aa}, which could allow lowering the background in \df\ searches \cite{1742-6596-718-6-062070}.
\jn\ and \skg\ will be the first projects that will have a substantial chance to observe the \df\ within  the next decade.

The purpose of the present paper is to offer an updated study of the \df, its sensitivity to failed \sne, and its detectability in the light of the latest theoretical and experimental advances. The \df\ is modeled using a recent set of detailed simulations of exploding and failed \sne\ from the Garching group \cite{phdthesis,Mirizzi:2015eza}. Recent theoretical results on what progenitor  stars are more likely to lead to \bh\ formation are incorporated as well.  We discuss the detectability of the \df\ at \skg\ and \jn, using current, detailed estimates of the relevant backgrounds. We consider the limitations posed by the low statistics at these detectors, and 
address the question of how likely it is that the \df\ will be discovered with high significance.

The paper is structured as follows.  In Sec. \ref{sec:formulation}, the basic physics of the \df\ is reviewed, and the inputs and assumptions of our calculation are described in detail.  The results for the \df\ are then presented in Sec. \ref{sec:flux}.  Sec. \ref{sec:detection} follows with a discussion of the flux detection potential at \skg, \jn\ and a possible slow liquid scintillator detector. Conclusions are then given in Sec. \ref{sec:discussion}.  Supplemental material is offered in three appendices.

\section{Formulation}
\label{sec:formulation}

In this Section, the essential elements of our calculation of the \df\ are given. We refer to Appendix \ref{A1} for details on the input \n\ flavor fluxes and luminosities.  

\subsection{Supernova progenitors and cosmological supernova rate}
\label{sub:progenitors}

The intensity and spectrum of the \df\ depend on the cosmological rate of core collapse (or, shortly, Supernova Rate, SNR).  
%
The SNR, differential in the progenitor mass $M$, $\dot{\rho}(z,M)$, is proportional to the star formation rate, $R_{SF}(z)$ (defined as the mass that forms stars per unit comoving volume per unit time, at redshift $z$): 
\begin{equation}\label{eq: 2.1}
\dot{\rho}(z,M)=R_{SF}(z) \frac{\phi(M)}{\int^{125M_{\odot}}_{0.5M_{\odot}}M\phi(M)dM}~,
\end{equation}
where $M_{\odot}=1.99\times10^{30}$ kg is the mass of the Sun, and  $\phi(M)$ is the Initial Mass Function (IMF), describing the mass distribution of stars at birth. 
The Salpeter IMF, $\phi(M) \propto M^{-2.35}$ \cite{1955ApJ...121..161S} is used here.

 The SFR is well described by the functional fit\footnote{Other proposed functional forms (see e.g., \cite{doi:10.1046/j.1365-8711.2001.04591.x, 2041-8205-802-2-L19}) give nearly identical results for the \df.  } \cite{0004-637X-651-1-142} \begin{equation}\label{eq: 2.2}
R_{SF}(z)= R_{SF}(0)\left\{
        \begin{array}{lll}
            (1+z)^{\beta} & \quad 0< z < 1 \\
            2^{\beta-\alpha}(1+z)^{\alpha}& \quad 1<  z < 4.5 \\
         2^{\beta-\alpha} 5.5^{\alpha-\gamma}(1+z)^{\gamma} & \quad 4.5<  z < 5 \\
        \end{array}
    \right.~,
\end{equation}    
where $\alpha=-0.26$, $\beta=3.28$, $\gamma=-7.8$, and $R_{SF}(0)=\mathcal{O}(10^{-2})~{\rm M_{\odot} Mpc^{-3} yr^{-1}}$.  Following ref. \cite{Lien:2010aa}, we take the total \snr\ normalization  to be
$R_{cc}(0)=\int^{125M_{\odot}}_{8M_{\odot}}\dot{\rho}(0,M) dM=(1.25\pm0.5)\times10^{-4}\rm{yr^{-1} }\rm{Mpc^{-3}}$.  

When addressing the question of what stars result in failed \sne,
one needs to consider that  the outcome of the collapse (explosion or direct black hole formation) depends on the interplay of several factors, and not directly on $M$. As a result, the distribution of \dbh\ with $M$ follows a complex pattern  \cite{0004-637X-757-1-69,0004-637X-762-2-126,0004-637X-801-2-90}. 

\begin{figure}[htbp]
\begin{center}
\includegraphics[width=0.7 \linewidth]{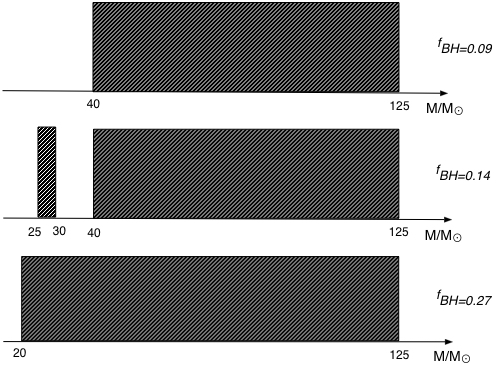}
\end{center}
\caption{\label{figure1}Different scenarios for the intervals of progenitor mass where direct black hole formation can be expected (shaded areas).  The legend shows the corresponding fractions of \dbh,  $f_{BH}$. }
\end{figure}

As examples, here we consider three possibilities, shown in Fig. \ref{figure1}. They are labeled by the fraction $f_{BH}$ of collapses that result in direct \bh\ formation, $f_{BH}=\int_{\Sigma}\phi(M)dM/\int^{125M_{\odot}}_{8M_{\odot}}\phi(M)dM$,
 where $\Sigma$ is the region of values of $M$ where BH formation is expected.
%

\begin{itemize}

\item In the first case, all stars with $M \geq 40 \msun$ result in failed supernovae,
 corresponding to $f_{BH}=0.09$.  This scenario appeared in early literature \cite{Woosley:2002aa}, and was used in the first studies of diffuse \ns\ from failed \sne\ (e.g., \cite{Lunardini:2009aa}). It represents the general situation where direct \bh\ production is rare, with only minor effects on the \df.

\item  The second case, considers \bh\ formation for $M \geq 40 \msun$, and also in the mass range $M=(25-30)\msun$.  The total \dbh\ fraction is  $f_{BH}=0.14$. The island of \bh\ formation at intermediate mass is mainly inspired by \cite{0004-637X-801-2-90} (``case a'' there, for solar-metallicity stars), and is also consistent with the results of \cite{0004-637X-757-1-69}.  The explanation of it is that stars with  $M\sim (20-30) \msun$ have high compactness, which is a characteristic of the density structure of the progenitor, and therefore are more likely to form black holes \cite{doi:10.1093/mnrasl/slu146,0004-637X-757-1-69}.

\item In the third scenario all stars with $M \geq 20 \msun$ collapse into a black hole, corresponding to $f_{BH}=0.27$. This case is similar to the scenario in \cite{0004-637X-827-1-85}, where the interval  $M \geq (16.5-18) \msun$ was considered for BH formation, and  was found to solve both the red supergiant problem \cite{doi:10.1146/annurev-astro-082708-101737} and the supernova rate problem \cite{0004-637X-827-1-85}.   
Moreover, a fraction of \dbh\ at the level of $\sim 30\%$ is suggested by the observation of a candidate failed \sn\ in a decade-long survey \cite{Adams:2016ffj}. Such fraction is well within observational constraints, $f_{BH}\lta$50$\%$ \cite{Adams:2016aa}. 

\end{itemize}

As a cautionary note, we stress that the mechanism of collapse into a black hole
is still not fully understood, therefore our results based on the scenarios above have the character of illustration only.

\subsection{Neutrino emission, propagation and flux at Earth}
\label{sub:fluxes}

For the neutrino flux from each supernova, we use the results of the spherically symmetric numerical simulation from the Garching group \cite{phdthesis,Mirizzi:2015eza}
for four different progenitor models (masses $M=11.2 M_{\odot}, 25 M_{\odot}, 27 M_{\odot}, 40M_{\odot}$) of solar metallicity, taken from Woosley and Weaver \cite{Woosley:2002aa}.
All the simulations use the Lattimer and Swesty equation of state \cite{LATTIMER1991331}, with compressibility parameter $K=220$ (LS220 from here on).
%
%
%
Protoneutron star formation is the outcome for the  $M=11.2 \msun, 27 M_{\odot}$ progenitors. For the case with $M=25 \msun$, two runs are used here, one for either outcome (\bh\ or NS).  The results for  $M=40M_{\odot}$ are for direct \bh\ formation. 

The numerical output files give the time-dependent luminosities and energy spectra of the three \n\ species $\nue$, $\barnue$ and $\nux$, where $\nux$ collectively denotes the non-electron flavors ($\numu, \nutau, \barnumu, \barnutau$). 
From these, the  time-integrated flavor spectra, $F^0_{w}$ ($w=e,\bar e, x$), are calculated.  These spectra exhibit the well known features of \dbh\ (see e. g. \cite{Sumiyoshi:2006aa}); one of them is the hotter spectra for all flavors, compared to \nsf.   In particular, the average energies for $\nue$ and $\barnue$ are higher by $\sim 3$ MeV, exceeding $\sim 18$ MeV for $\barnue$.  The failed \sn\ \n\ flux tends to deviate from the flavor equipartition of energy (which characterizes exploding \sne), in favor of the electron flavors: for $M=40 \msun$, the luminosity in $\nux$ is almost half of the one in $\nue$. 
A full description of the flavor spectra $F^0_{w}$ is given in the Appendix \ref{A1} (see Table \ref{Table:1} and Fig. \ref{figure2} there). 

%
%

To fix the ideas, let us consider the $\barnue$ component of the \n\ emission, which is most relevant for detection (Sec. \ref{sec:detection}). 
The $\bar{\nu}_e$ spectrum reaching a detector on Earth is different than the $\bar{\nu}_e$ spectrum at  production, due to neutrino oscillation inside the supernova envelope \cite{PhysRevD.62.033007}.  After oscillations, the $\barnue$ spectrum can be written, in terms of the time-integrated flavor spectra $F^0_{w}$, as: 
\begin{equation}\label{eq: 2.6}
F_{\bar{\nu}_e}= \bar{p} F^0_{\bar{\nu}_e} +(1-\bar{p}) F^0_{\bar{\nu}_x}~,
\end{equation}
where $\bar{p}$ is a energy-dependent probability describing the amount of flavor permutation. This quantity is difficult to estimate, due to the effect of collective oscillations near the neutrinosphere (see e.g., \cite{1475-7516-2012-07-012}), which is only partially understood. 
For illustration, here results will be shown for  $\bar{p}=0$ and $\bar{p}=0.68$. These are the extreme values that $\bar{p}$ can take with the assumption that the lower density Mikheev-Smirnov-Wolfenstein (MSW) resonance \cite{PhysRevD.17.2369,0038-5670-30-9-R01} is adiabatic and decoupled from earlier oscillation stages (i.e, collective oscillations and higher density MSW resonance). 
It is possible that such extreme values might be realistic: indeed, preliminary studies found that if collective oscillations are active only in the cooling phase of the burst, and suppressed in the accretion phase, their effect on the time-integrated \n\ flavor spectra might be less than $\sim 10\%$ \cite{1475-7516-2012-07-012}. Therefore, the  value of $\bar p$ would be mostly due to the MSW effects, with  $\bar{p}\simeq 0.68$ ($\bar{p}\simeq 0$) for the normal (inverted) neutrino mass hierarchy \footnote{Rigorously, for the inverted mass hierarchy, MSW resonant flavor conversion predicts $\bar p= \sin^2 \theta_{13} \simeq 2 \times 10^{-2}$ \cite{PhysRevD.62.033007}. Here the small contribution of this term is neglected.}.


The diffuse flux of $\barnue$ in a detector at the Earth, differential in energy and surface is given by  \cite{1367-2630-6-1-170}:
\begin{equation}\label{eq: 3.1}
\Phi(E)=\frac{c}{H_0}\int_{8  M_{\odot}}^{125 M_{\odot}}\int^{z_{max}}_0 \dot{\rho}(z,M)  \frac{dF_{\bar{e}}(E(1+z),M)}{dM} \frac{dz}{\sqrt{\Omega_m(1+z)^3+\Omega_{\Lambda}}}dM
\end{equation}
where $\Omega_m=0.3$, $\Omega_{\Lambda}=0.7$ are the fractions of the cosmic energy density in matter and dark energy; c is the speed of light, $H_0$ is the Hubble constant. Here $dF_{\bar{e}}(E(1+z),M)/dM$ is the number of $\barnue$ per unit energy (after oscillations) produced by an individual \sn\ with progenitor mass between $M$ and $M+dM$.  

Since numerical results are available only for discrete values of $M$, the dependence of  $dF_{\bar{e}}(E(1+z),M)/dM$ on $M$ was approximated as a step function, i.e., as a constant in $M$ in certain given mass intervals $[ M_i , M_{i+1} ]$, as done in \cite{1475-7516-2012-07-012}. The intervals are selected to reproduce the three cases in fig. \ref{figure1}; and for each interval, the numerical run 
with $M$ such that $ M_i < M <M_{i+1} $ is taken as representative of the entire interval.  

In Eq. (\ref{eq: 3.1}), $z_{max}=2$ was chosen. We excluded the interval $z > 2$ because the \n\ fluxes we use are for solar metallicity, and therefore they become increasingly inaccurate for increasing $z$ (which corresponds to a decrease in the metallicity of stars).  Therefore, in this work the \df\ is underestimated. The error is potentially large at $E \lta 8$ MeV or so,  but our calculations show that it is likely to be negligible above realistic detection thresholds, $E \gta 11$ MeV. 


\section{The diffuse flux}\label{sec:flux}

Here our results for the \df\ are discussed. They are summarized in Table \ref{Table:2} and Fig. \ref{figure5}.  More details on the dependence of the \df\ spectrum on the parameters can be found in Appendix \ref{B}. 

To illustrate the interval of values that can be expected for the flux, we varied the normalization of the total rate of core collapses in the interval $R_{cc}(0)=(1.25 \pm 0.5)\times10^{-4}\rm{yr^{-1} Mpc^{-3}}$ \cite{Lien:2010aa}. 
%
 %
Results are presented for three representative scenarios, where the central and extreme values of $R_{cc}$ are combined with the cases in Fig. \ref{figure1} (for $\bar p=0.68$) to give moderate, suppressed or enhanced \n\ flux. The scenarios will be called Low, Fiducial and High (corresponding to the intensity of the \df); they are detailed in Table \ref{Table:2}, for $\bar p=0.68$, and in two bins of neutrino energy, $[11, 30]~{\rm MeV}$ and  $[30, 50]~{\rm MeV}$.
  For each case and each bin, the Table gives the flux (total and \dbh\ only) and a spectrum parameter, $\epsilon_0$, defined as the energy for which an exponential form $\Phi(E) \propto e^{-E/\epsilon_0}$, fits the spectrum best in the energy interval of interest\footnote{In general, the spectrum resembles the superposition of two exponential spectra, one for \nsf\ and the other for \dbh\ (see Appendix \ref{B}).  However, we find that locally, for energy bins of width $\Delta E \lta 20$ MeV, a single exponential form is still an adequate approximation (at the level of $\sim 1\%$ or less in the respective energy window) that can be useful in future data analyses.}. 
 
  The Table shows that the flux at $E \gta 11$ MeV can be as large as $\phi=3.7~{\rm cm^{-2} s^{-1}}$.  Expectedly, the contribution of \dbh\ increases from a modest $\sim 20\%$ in the lower energy bin, to $\sim 70\%$ in the higher energy bin, depending on the parameters.   This corresponds to a change in  $\epsilon_0$, which can be as large as  $\epsilon_0\simeq 6$ MeV in the higher energy interval.

An important point to notice is that in the lower energy bin, $\phi$ depends very little on the scenario of black hole formation (Fig. \ref{figure1}): when changing from the Low to Fiducial and High cases, most of the flux increase 
  is due to the increase of the \snr\ normalization, with only minor variations due to the change in the distribution of failed \sne\ with the progenitor mass. 
%
This can be understood considering that in the intermediate mass region, $M\simeq 25\msun$, the \n\ emission for \nsf\ and \dbh\ is overall similar in the $E \sim 10-20$ MeV energy interval (see Appendix \ref{A1}). 
This feature may depend on the details of the numerical simulations used here. If it is confirmed to be robust, it may allow to use the lower part of the \df\ energy spectrum to test the normalization of the core collapse rate independently of the specific pattern of \bh\ formation; while the higher part of the energy spectrum could, in principle, reveal the effect of failed supernovae in the increase of $\epsilon_0$.

  \begin{table}[htbp]
\centering
\begin{tabular}{|c| c| c| c| c|}
\hline\hline
Energy window&Parameters & Low&Fiducial &High \\ [0.5ex]  
(MeV)&&&&\\
&&&&\\
\hline
 &&&&\\
 &$f_{BH}$ \tablefootnote{We stress that here $f_{BH}$ is not an input parameter of the calculation, but rather a useful label of the different possibilities considered here (Fig. \ref{figure1}). For each of them, the \df\ is calculated including the  progenitor-dependence of the neutrino fluxes, as discussed earlier in this Section.}
&0.09 & 0.14&0.27 \\
 &&&&\\
\hline
 &&&&\\
 &$R_{CC}(0)(10^{-4}\rm{yr^{-1} }\rm{Mpc^{-3}}$)& 0.75& 1.25&1.75\\
&&&&\\
\hline
 &&&&\\
\multirow{4}{1cm}{11-30}&$\phi$(cm$^{-2}$ s$^{-1}$)&1.5 [0.28] & 2.5 [0.65]&3.6 [1.63] \\
&&&&\\
&$\epsilon_0$ (MeV)  &4.77 & 4.86&5.12 \\
&&&&\\
\hline
 &&&&\\
\multirow{4}{1cm}{30-50}&$\phi$(cm$^{-2}$ s$^{-1}$)& 0.03 [0.015] & 0.06 [0.03]&0.11 [0.08] \\
&&&&\\
&$\epsilon_0$ (MeV) & 5.67 & 5.81&6.1 \\
 &&&&\\
\hline\hline
 \end{tabular}
\caption{Results for three different combinations of \snr\ normalization and fraction of \dbh\ (as in Fig. \ref{figure1}). Two energy bins are considered, and for each bin we give the flux and the energy parameter $\epsilon_0$ that appears in an exponential approximation of the spectrum.  Numbers written in square brackets correspond to the contribution from the BHFCs.  $\bar p=0.68$ was used here.  }  \label{Table:2}\end{table}

\begin{figure}[htbp]
\begin{center}
\includegraphics[width=0.65 \textwidth]{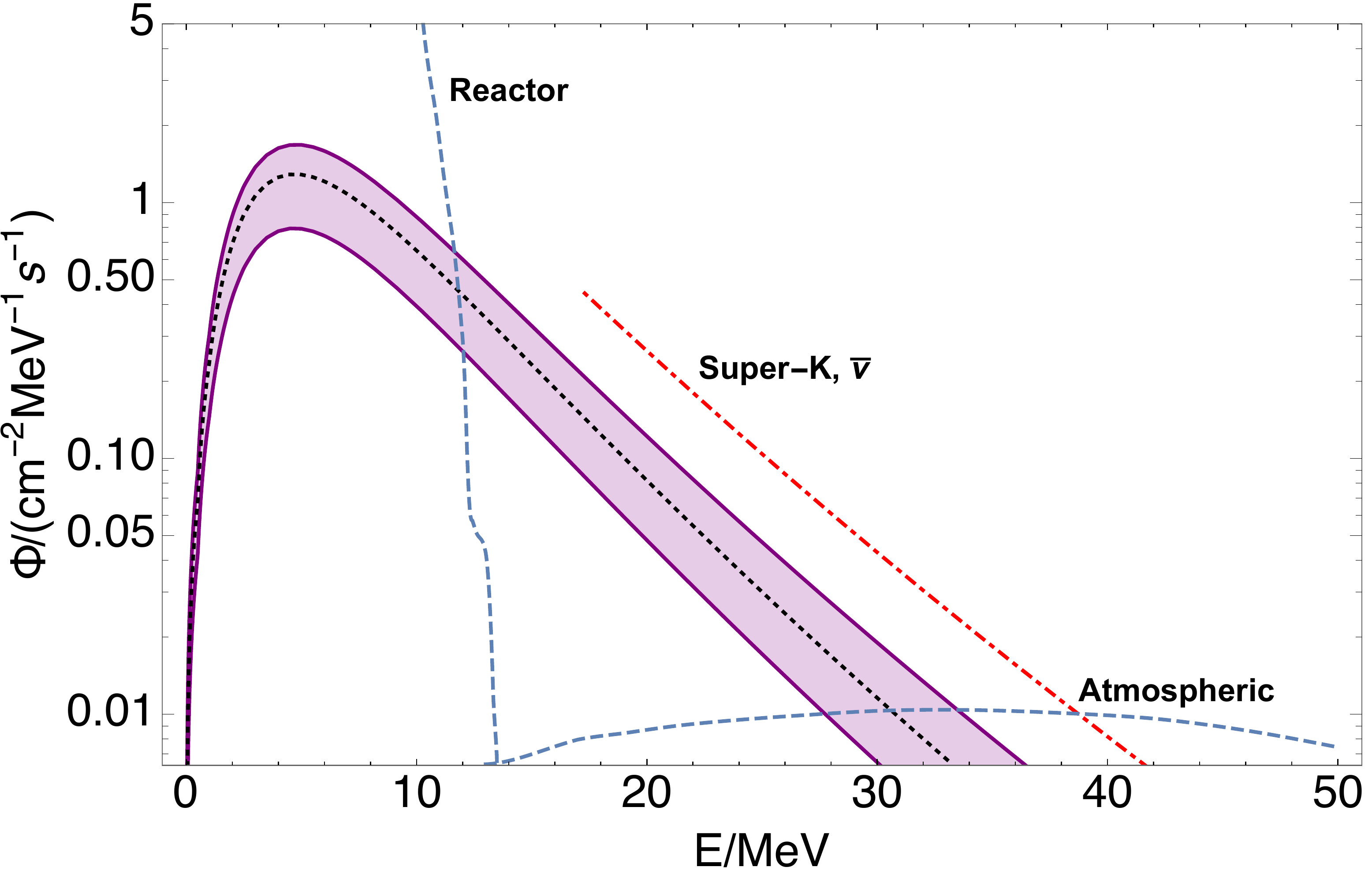}
\end{center}
\caption{\label{figure5} Summary: the Diffuse Supernova Neutrino Background (total of neutron-star-forming- and black-hole-forming- collapses), for $\bar p=0.68$, with uncertainties due to astrophysical inputs and to the fraction of black hole-forming collapses (shaded area).  Shown is the predicted flux for the Fiducial (dotted line), Low and High cases (solid lines), as in Table \ref{Table:2}. 
  Background $\barnue$ fluxes are shown as dashed lines:  from nuclear reactors at lower energy (taken from  \cite{PhysRevD.75.023007}) and from the atmosphere at higher energy \cite{Battistoni2005526}, for the Kamioka site.
  For comparison, we also show a signal flux (dot-dashed line, same spectrum as the Fiducial case) that would saturate the current \sk\ upper bound \cite{PhysRevD.85.052007} (see text). }
\end{figure}

In Fig. \ref{figure5} we show the range of diffuse flux spanned by the three cases in table \ref{Table:2}. The background fluxes from reactor and atmospheric neutrinos at the Kamioka site are also shown in the figure. The reactor antineutrino flux is a serious hindrance to the study of the \df; for example, at the Kamioka site the reactor flux is as high as $\Phi_{re} \simeq 10^2$ cm$^{-2}$ s$^{-1}$MeV$^{-1}$ at 8 MeV, and dominates over the DSNB below 12 MeV. The atmospheric neutrino background exceeds the DSNB above $\sim$ 31 MeV or so depending on the intensity and spectrum of the DSNB. Therefore, here the energy window of experimental interest is set to be $\sim$ 11-30 MeV. 

 It is interesting to compare our predicted $\barnue$ flux with the Super-Kamiokande bound above $17.3$ MeV ($\barnue$ energy): 
 $ \phi(E \geq 17.3~{\rm MeV})\leq 3.0 ~{\rm cm^{-2} s^{-1}}$ at 90\% C.L.  \cite{PhysRevD.85.052007}.
 Fig. \ref{figure5} shows the \df\ (with the same spectrum as the Fiducial case), with the normalization increased so to saturate the bound. The flux for the High case is a factor $\sim 3$ below the \sk\ bound, therefore an improvement of at least a factor of a few is needed in the experimental reach to achieve detection.  


\section{Detection and discovery potential}\label{sec:detection}

Let us consider the detection of $\barnue$ via inverse beta decay (IBD, $\barnue+p\rightarrow n+e^+$), which is the dominant detection process in both water \ck\ and liquid scintillator detectors. 
The minimum \n\ energy needed to initiate the IBD reaction (Q value) is 1.806 MeV. 
Thus, the positron kinetic energy is  
 \begin{equation} \label{eq: 4.2}
 K_{e^+}=E_{\nu}-1.806~{\rm MeV}~. 
 \end{equation}
 The differential rate of detection (number of events per unit energy per unit time) of the \df\ is 
 \begin{equation}
\frac{dN}{dE_{\nu}}=f_{eff} \Phi(E_\nu) \sigma_{\nu}(E_{\nu}) N_p~,
\label{eq:nevents}
\end{equation}
 where $f_{eff}\leq 1$ is the detector efficiency, N$_p$ is the number of protons in the target volume, and $\sigma_{\nu}(E_\nu)$ is the cross section of the IBD reaction. Here the cross section  from \cite{Strumia200342} will be used.  

\subsection{Detectors and backgrounds}
\label{subsec:detectors}

\subsubsection{Liquid Scintillator}
\label{subsubsec:lab}

Motivated by the upcoming JUNO detector, we consider a multi-kiloton liquid scintillator (LS) detector, employing LAB (Linear Alkyl Benzene) as a realistic candidate target material. 
The positron produced from the IBD reaction annihilates with an ambient electron,  producing visible light with energy (from Eq. (\ref{eq: 4.2})): 
 \begin{equation}
 E_{vis}=E_{\nu}-0.8~{\rm MeV}~. 
 \end{equation}
A measurement of $E_{vis}$ then immediately gives the energy of the incoming neutrino.
An additional signature is the capture of the neutron on a free proton; a process that occurs with a delay $\tau_n \simeq 200$ $\mu$s (the average lifetime of the neutron in LAB), and produces a 2.2 MeV photon.
Thus, a  prompt-delayed coincident measurement of the IBD event is generally performed in LS, which greatly enhances the tagging power. 

The main backgrounds to the \df\ signal, in the energy window of experimental interest, are due to atmospheric neutrinos, mainly via charged current (CC) $\barnue$ scattering  and neutral current (NC) interactions. 
 Unlike in water Cherenkov detectors (Sec. \ref{subsec:water}), the background originating from CC atmospheric $\nu_\mu$'s and $\bar{\nu}_\mu$'s is less problematic in liquid scintillator since the final state muons can be tagged efficiently by  daughter electrons, and by the characteristic pulse shape of muon events \cite{0954-3899-43-3-030401}. Thus, this background can be neglected in our analysis. Likewise, the atmospheric $\nu_e$ CC events can be neglected, as they can be identified by a neutron tagging technique. 
The atmospheric neutrino NC events can produce an IBD-like signature; one of these is the ejection of a neutron from the carbon nucleus, with the nucleus being left in an excited state with multiple decay modes.  More complicated processes are also possible \cite{0004-637X-745-2-193}.  Most of the decays in such reactions occur over timescales that are much longer than $\tau_n$,  which allows  a $\sim$40\% rejection of the NC background \cite{Mollenberg:2015aa}.

In this work, two different detector configurations, with different levels of backgrounds, will be discussed:

\begin{itemize}

\item the setup envisioned for JUNO \cite{0954-3899-43-3-030401}, where the NC background in LAB can be reduced using  pulse shape discrimination  \cite{Mollenberg:2015aa}. This can be done at the cost of a decreased  signal efficiency, which is estimated to be $f_{eff} \simeq 50\%$ \cite{0954-3899-43-3-030401}. Here the detailed energy spectrum for the total residual background, as obtained in \cite{0954-3899-43-3-030401}, will be used. 
 
  \item  the technique 
proposed by Wei et al.,   \cite{Wei2017255}, who discuss  the use of LAB as a slow liquid scintillator (SLS from here on) in the context of  a possible future kt-scale detector \cite{1674-1137-41-2-023002}.   In SLS, it is possible  to separate the Cherenkov and scintillation lights \cite{Li2016303}, which allows to substantially reduce the atmospheric NC background, while maintaining high signal efficiency,  $f_{eff} \simeq 90\%$ \cite{Wei2017255}.  
The energy spectrum for the residual background in this case has been taken from fig. 4 in \cite{Wei2017255}.  For the sake of comparison, we will show results for SLS for the same exposure as JUNO. Such a large exposure is in principle possible, and has been suggested recently for other advanced liquid scintillator concepts \cite{Gann:2015aa}.
     
\end{itemize}   
  
\subsubsection{Water Cherenkov with Gadolinium}
\label{subsec:water}
The \skg\ experiment will be created
 by dissolving  gadolinium sulphate (Gd$_2$SO$_4$)  in the water of \sk\ in $\sim$ 0.2\%  concentration. This setup will allow tagging a IBD event by the capture of  the final state neutron on Gd, with an efficiency of $\sim$ 90\% \cite{Beacom:2004aa}. 
 The energy of the parent \n\ will be obtained from the measured (total) energy of the  positron,  $E_{e^+}=E_{\nu}-1.3~{\rm MeV}$, Eq. (\ref {eq: 4.2}).  

Several processes contribute to the background of the DSNB search in \skg.  Similarly to the case of LAB, reactor $\barnue$s represent an unsurmountable background, and determine the lower end of the energy window to be around $E_{e^+}\sim 11$ MeV.  Above this energy, the most important backgrounds are due to atmospheric neutrinos interactions, and in particular, to (i) $\bar{\nu}_{e}$ scattering (IBD) (ii) CC scattering of $\nu_{\mu}$/$\bar{\nu}_{\mu}$  (iii) NC elastic scattering and (iv) neutral current inelastic scattering with one pion production (NC1$\pi $). 
  Let us discuss these processes in order. 
 
 IBD events due to atmospheric $\barnue$ are indistinguishable from the signal, and therefore can not be reduced. As discussed before, they close the energy window from above at $E_{e^+ }\sim 30-40$ MeV.  
 $\nu_{\mu}$/$\bar{\nu}_{\mu}$ CC scattering can produce sub-Cherenkov $\mu^{\pm}$ (``invisible muons'') \cite{PhysRevLett.90.061101}, the decay of which mimics  the IBD reaction.  Due to the IBD tagging by Gd, the number of invisible muon decay events in the final sample (after cuts) can be reduced by a factor of $\sim 5$ \cite{Beacom:2004aa}.

The atmospheric NC elastic events lead to neutron knock-out off an oxygen nucleus, which then produces de-excitation photons \cite{PhysRevD.88.093004}. Similarly, the pions produced via NC1$\pi$ reactions are absorbed by oxygen nuclei in water, thus producing de-excitation $\gamma$ rays.   In both cases, the final state can mimic the IBD signature. These IBD-impostors can be excluded in part by Cherenkov angle selection cut ($\theta_c \approx$ 38-50 degrees), but some still leak into the final sample. Their energy spectrum rises sharply with decreasing energy. Therefore, with the lowering of the energy threshold due to the addition of gadolinium, the NC atmospheric \n\ \cite{phdthesis1,PhysRevD.85.052007}
 background has become much more relevant and needs to be modeled in detail. 

Here we use the backgrounds from the recent analysis in \cite{phdthesis1} (fig. 8.5 therein)\footnote{We note that the \df\ signal used in \cite{phdthesis1} is much larger (by a factor of $\sim$2-3) than the results of most literature (e.g., \cite{ANDO2003307,YUKSEL2015413,0004-637X-804-1-75}), thus leading to more optimistic conclusions about detectability than the present work.}.
We assume the signal efficiency of the detector to be about $67\%$ \cite{phdthesis1, Abe:2011aa}. 

 \subsection{Number of events}
 \label{subsec:eventrates}

\begin{figure}[htbp]
\begin{center}
\includegraphics[width=0.5 \textwidth]{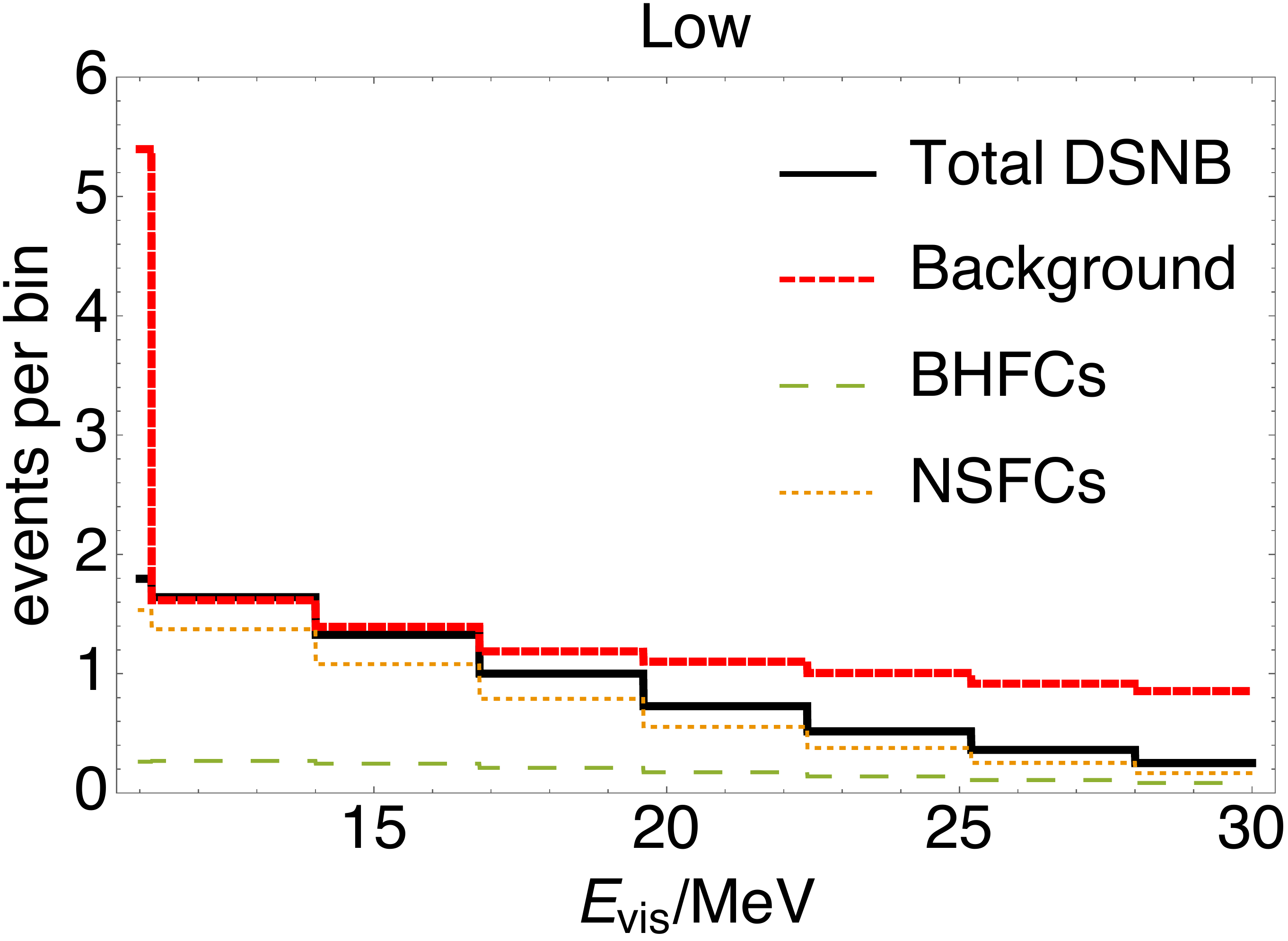}
\vskip0.4truecm
\includegraphics[width=0.5 \textwidth]{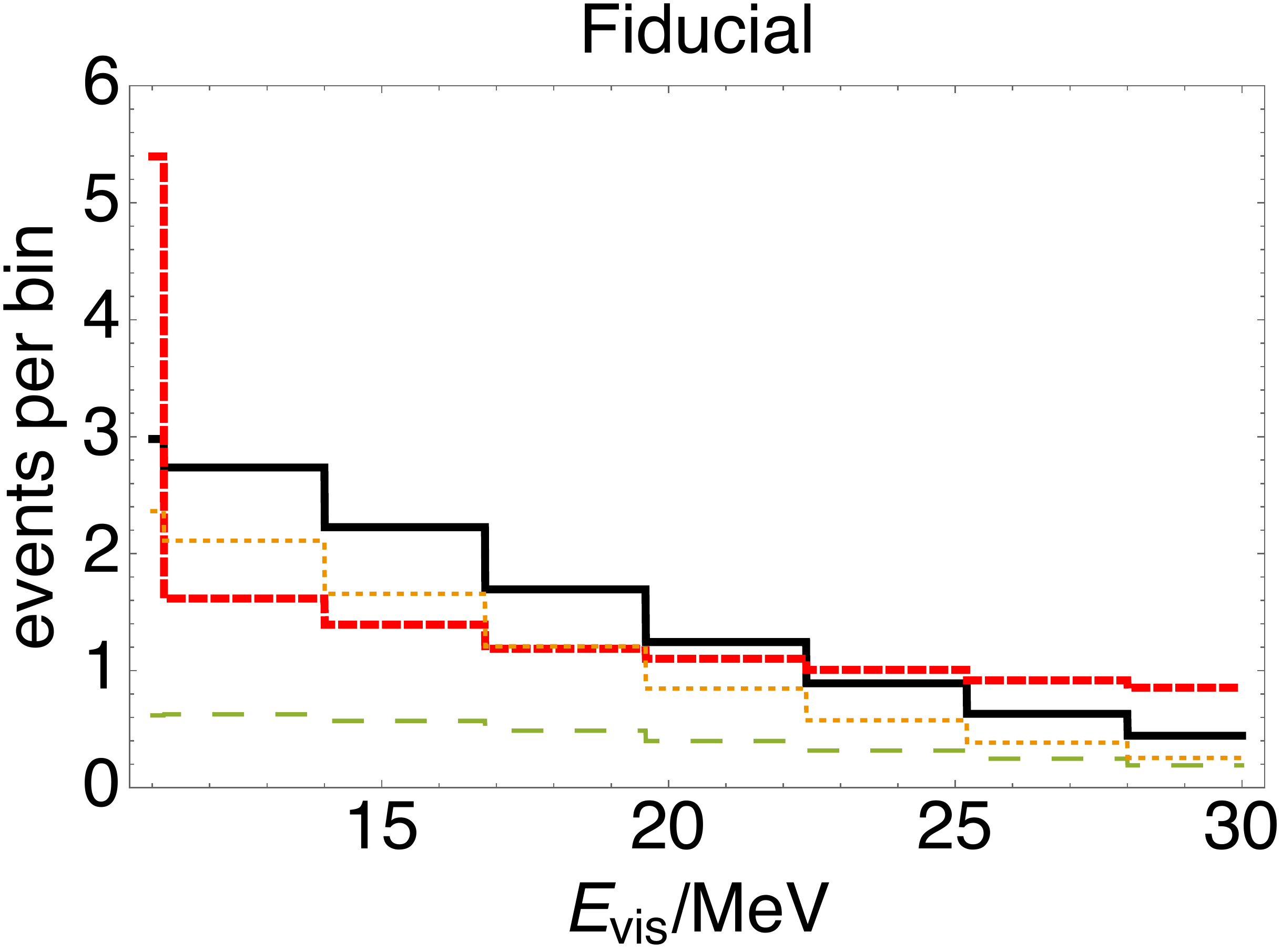}
\vskip0.4truecm
\includegraphics[width=0.5 \textwidth]{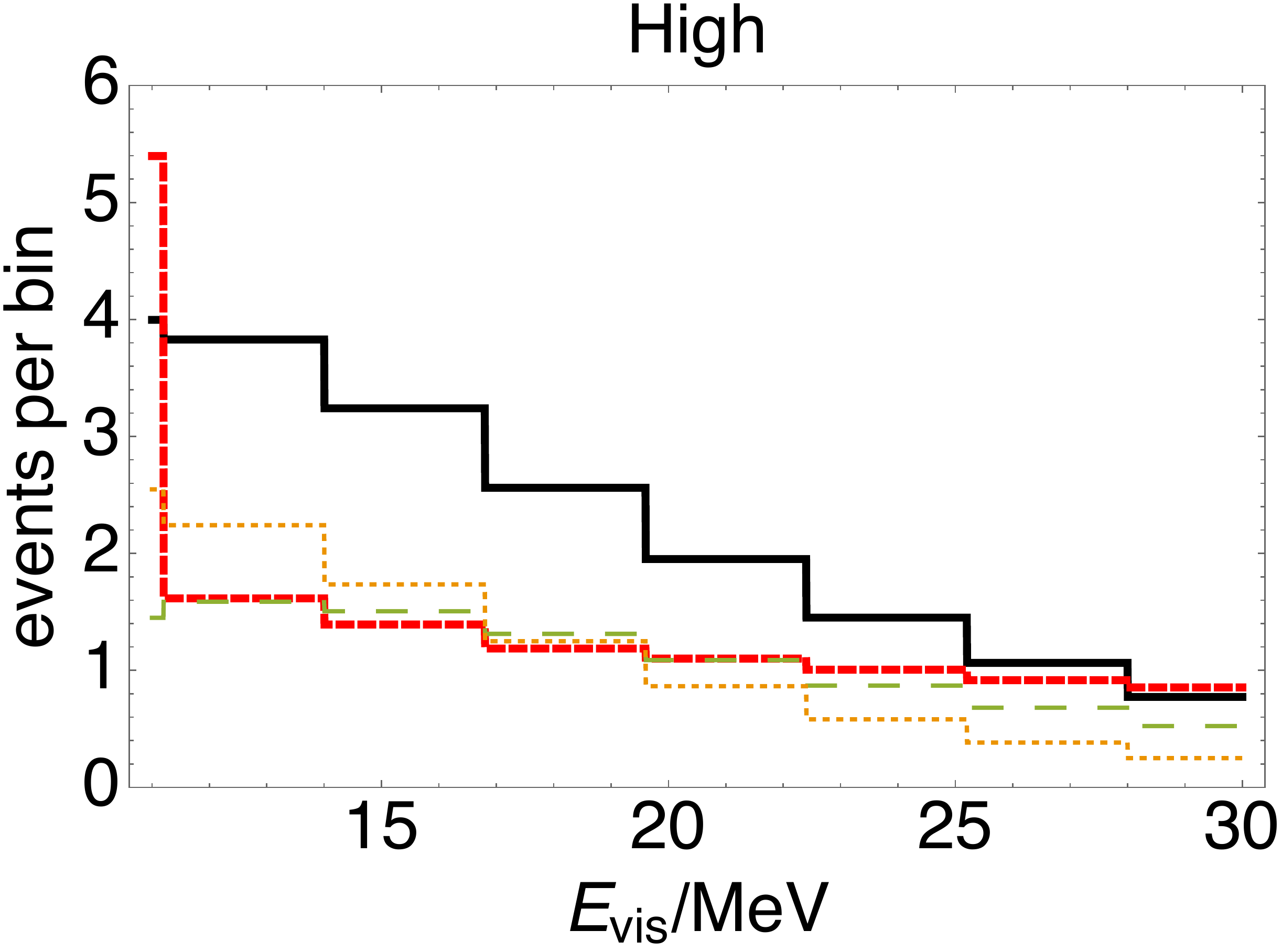}
\end{center}
\hspace{\fill}
\caption{\label{figure6} Number of \df\ (total, and contributions of \nsf\ and \dbh) and background events in JUNO, with an exposure ${\mathcal M}=200~{\rm kt ~yr}$, for the Low, Fiducial and High signal case (see Table \ref{Table:2}), for $\bar p=0.68$.}
\end{figure}
 
 \begin{figure}[htbp]
 \begin{center}
\includegraphics[width=0.5 \textwidth]{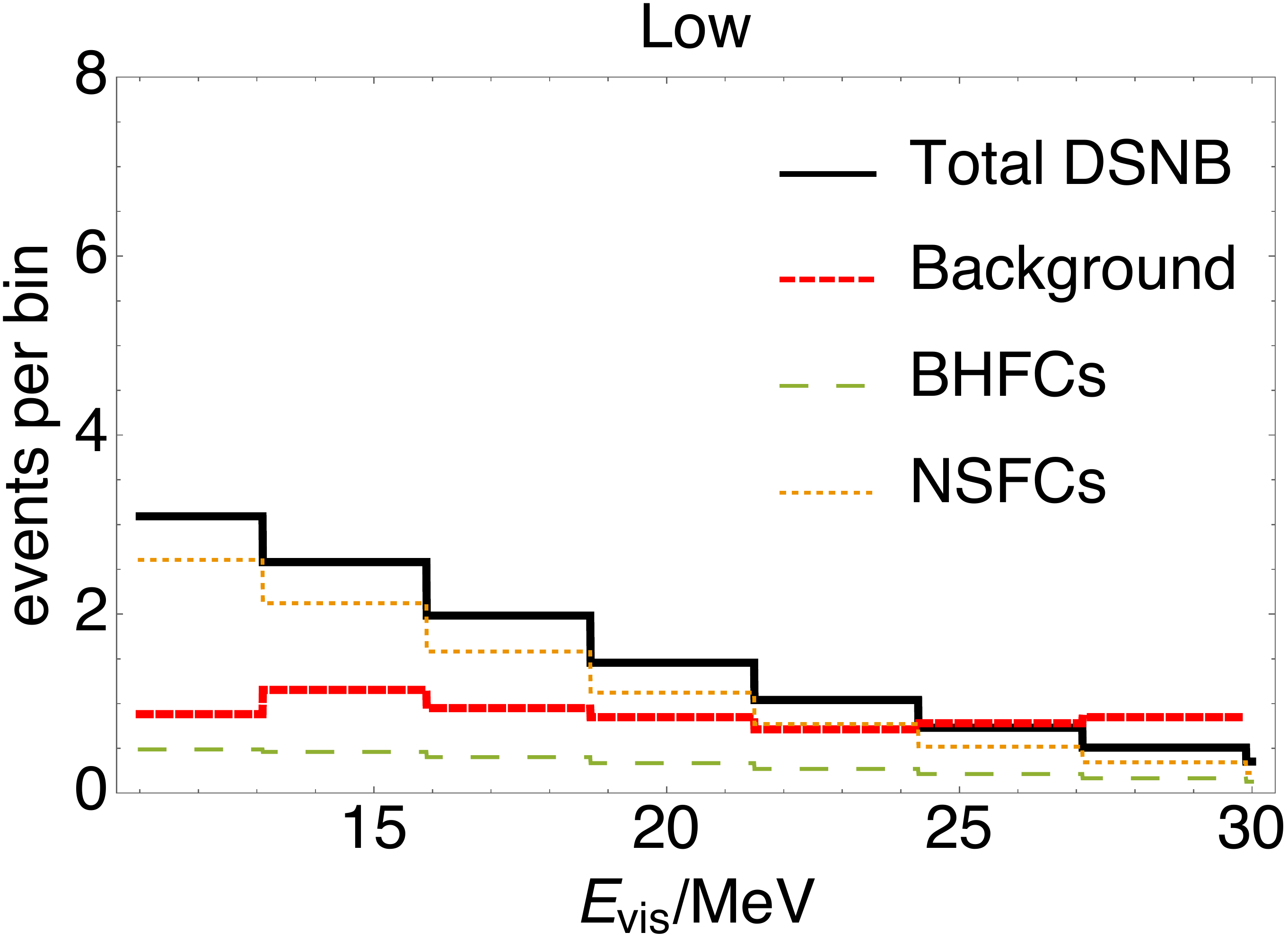}\\
\vskip0.4truecm
\includegraphics[width=0.5 \textwidth]{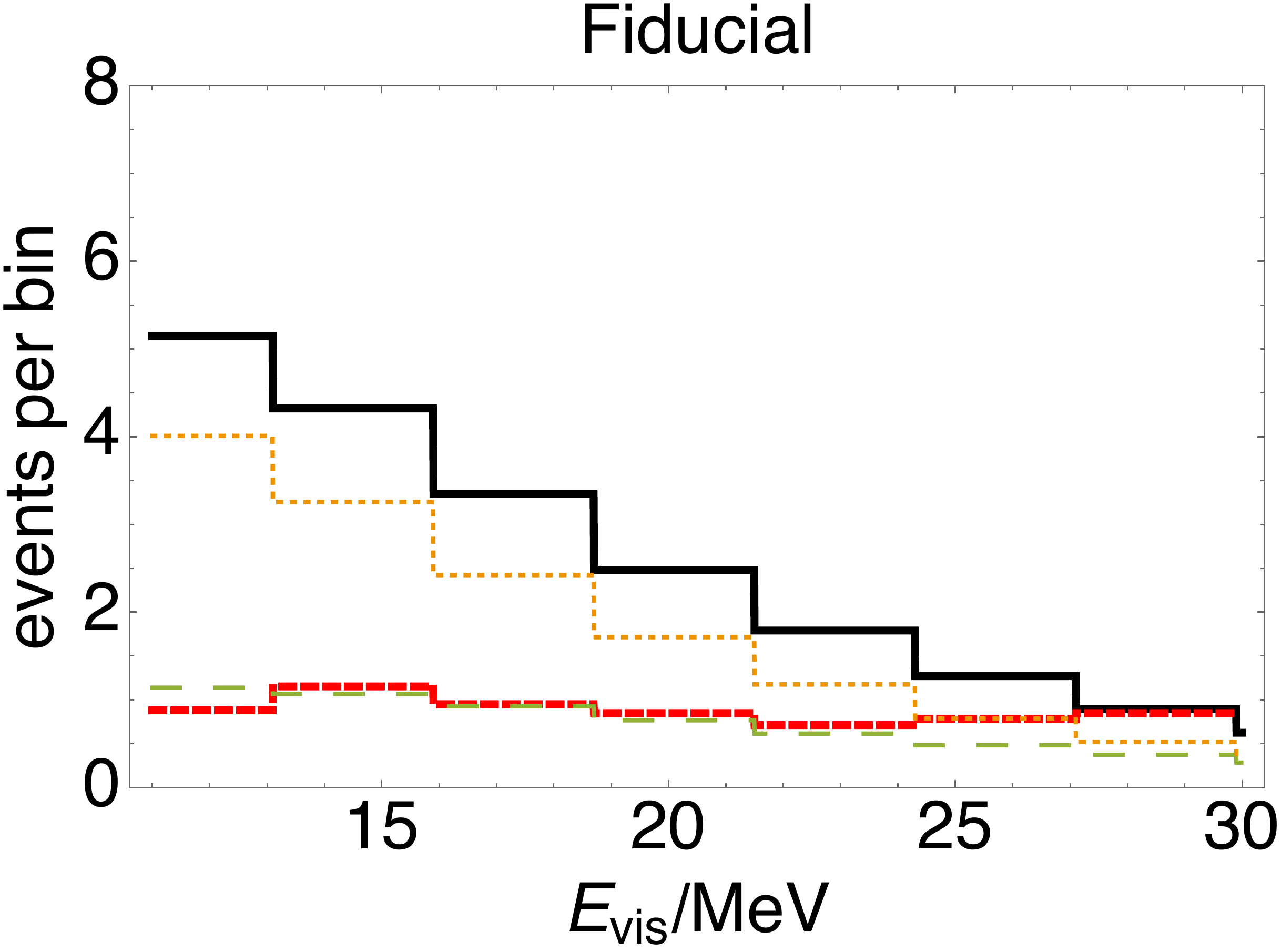}\\
\vskip0.4truecm
\includegraphics[width=0.5 \textwidth]{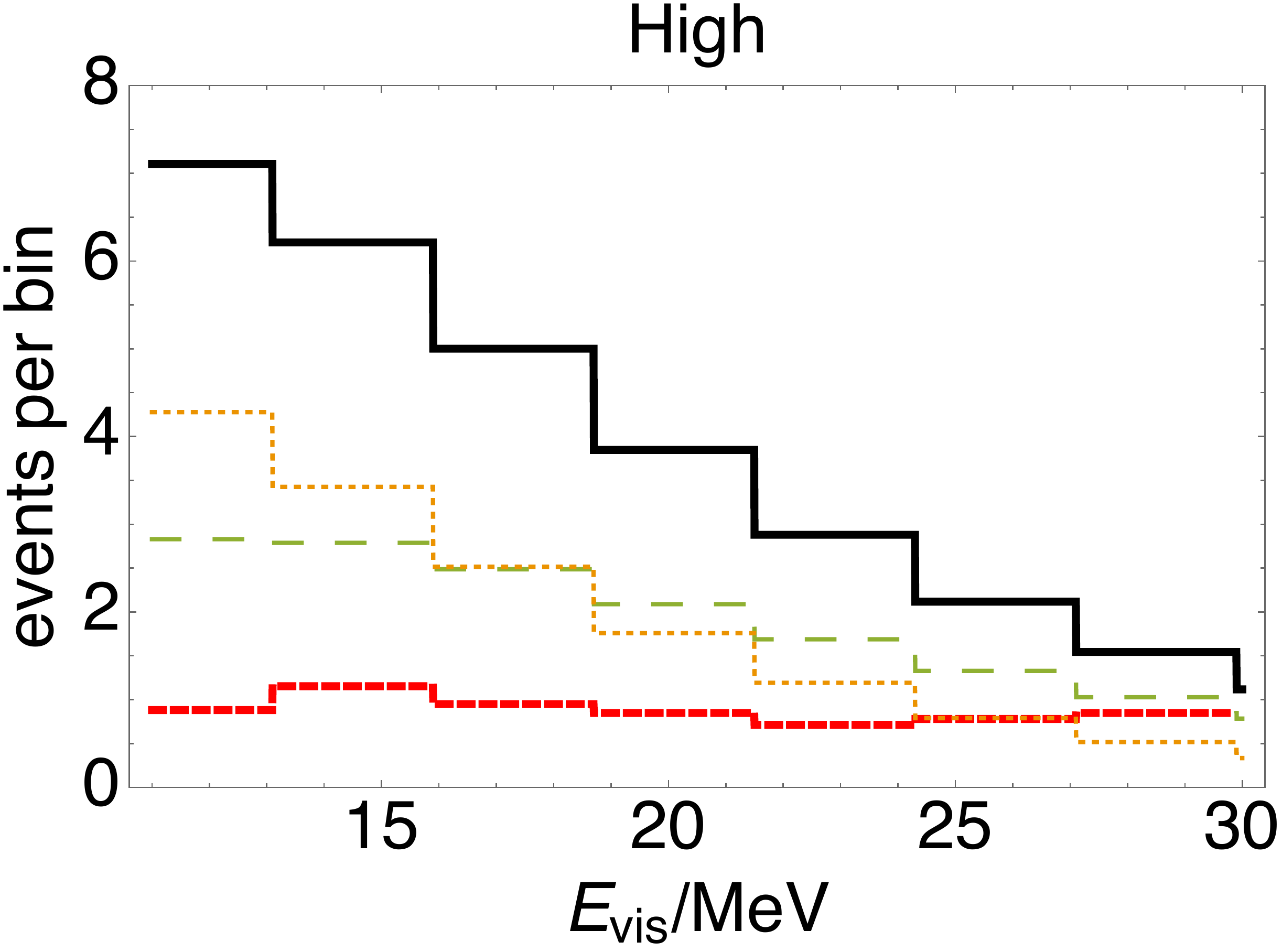}
\end{center}
\hspace{\fill}
\caption{\label{figure7} The same as fig. \ref{figure6} for SLS.  }

\end{figure}

 \begin{figure}[htbp]
 \begin{center}

\includegraphics[width=0.5 \textwidth]{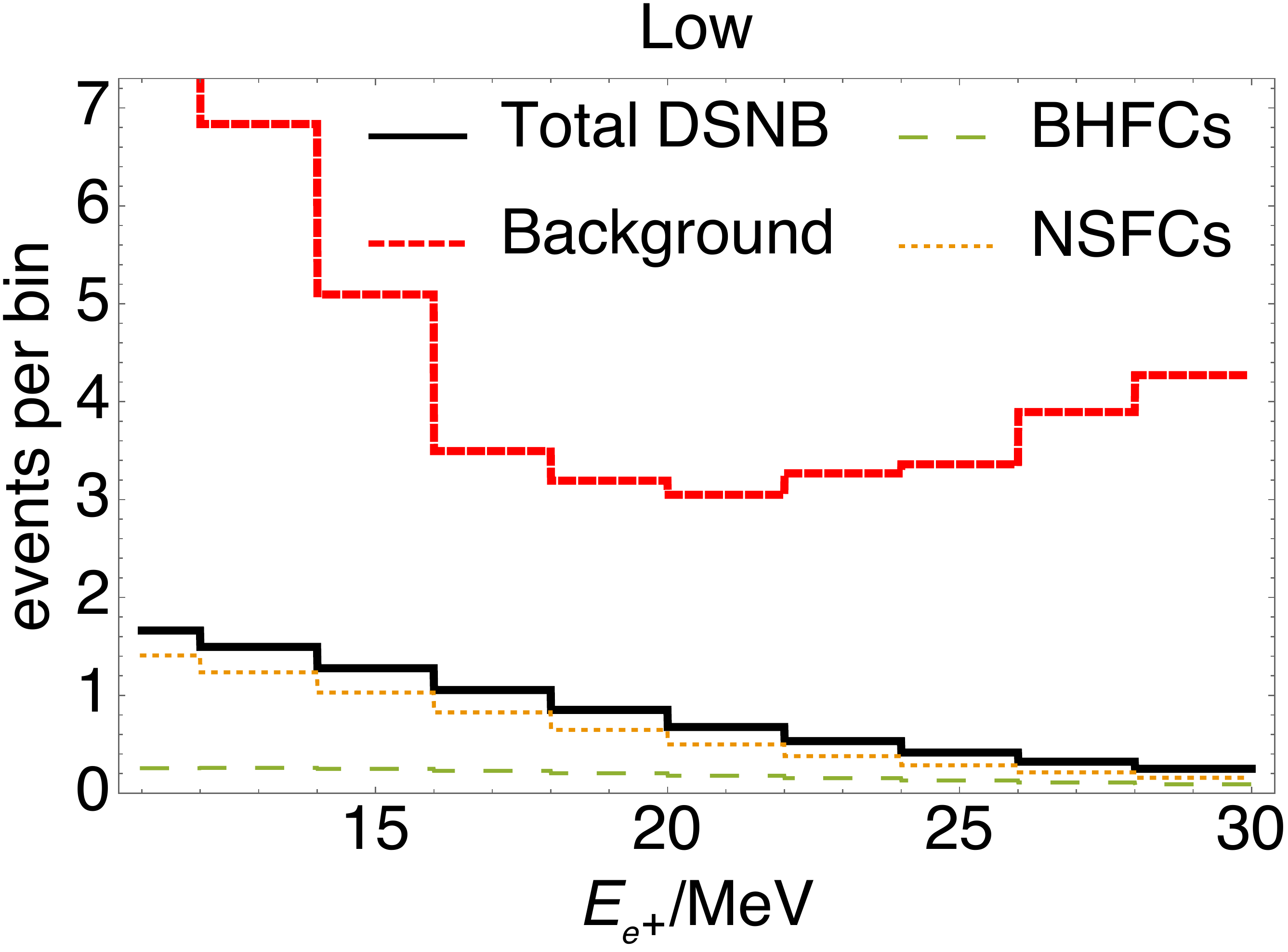}
\vskip0.4truecm
\includegraphics[width=0.5 \textwidth]{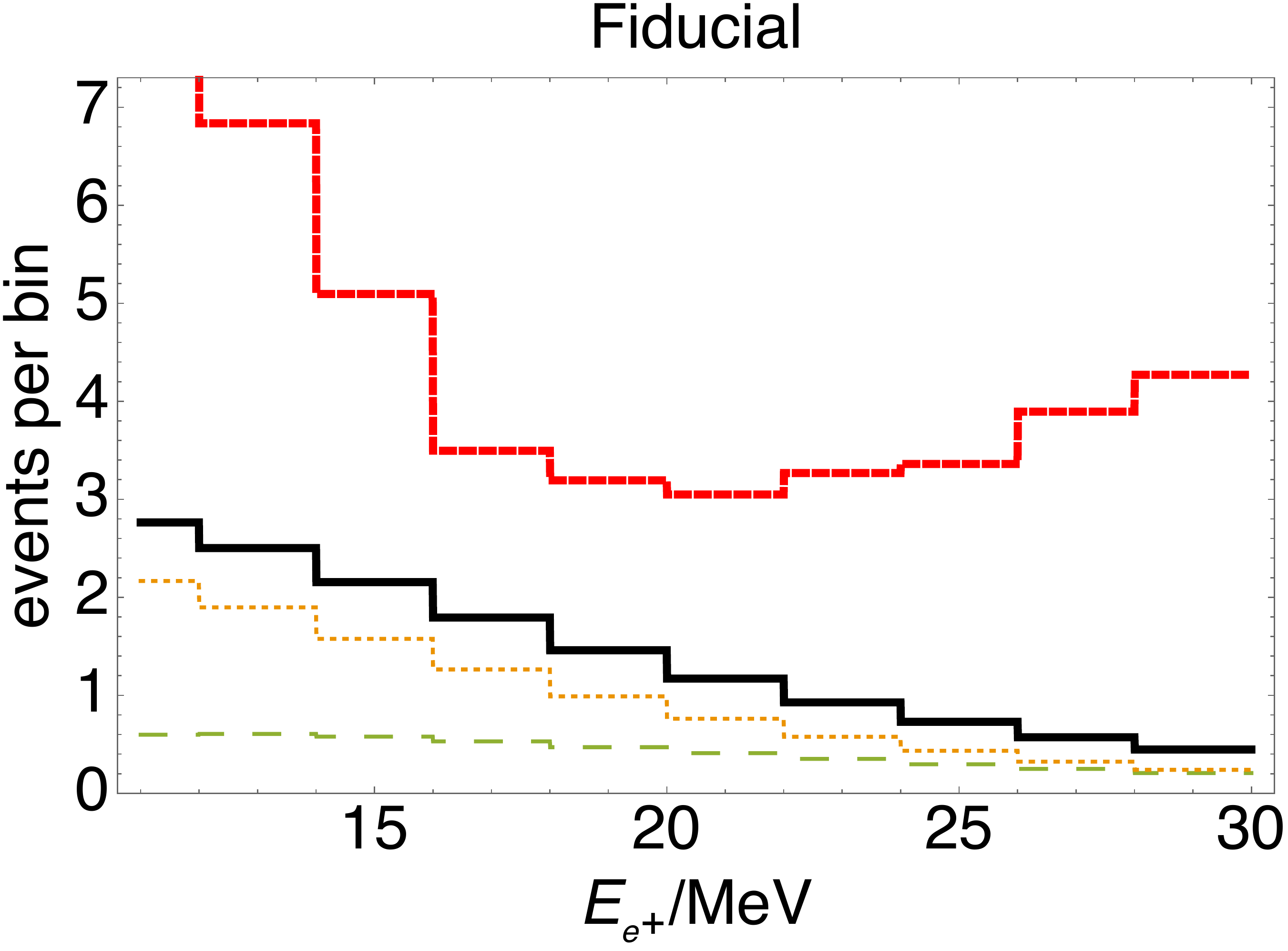}
\vskip0.4truecm
\includegraphics[width=0.5 \textwidth]{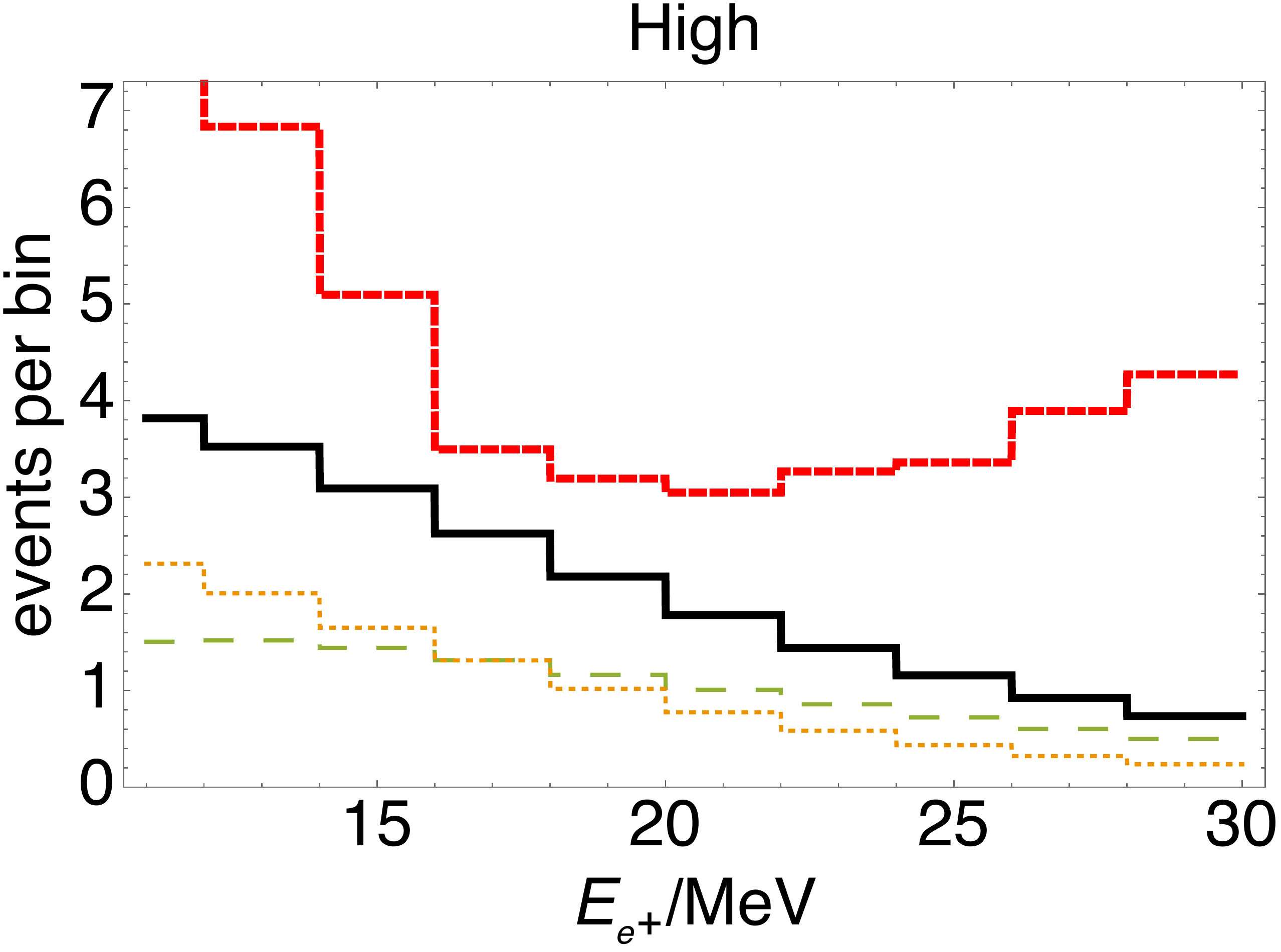}
\end{center}
\hspace{\fill}
\caption{\label{figure8} The same as fig. \ref{figure6} for \skg. Here the exposure is  ${\mathcal M}=225~{\rm kt~ yr}$.  }

\end{figure}

We calculated the number of signal and background events expected in the realistic energy window, for the three experimental setups of interest, and approximately $\Delta T=$10 years of data taking.  For brevity, results are shown only for $\bar p=0.68$.  They are summarized in Table \ref{table:3}.  They are also shown in Fig. \ref{figure6} for JUNO (exposure ${\mathcal M}=200~{\rm kt ~yr}$), Fig. \ref{figure7} for SLS (${\mathcal M}=200~{\rm kt ~ yr}$) and Fig. \ref{figure8}, for \skg\ (${\mathcal M}=225~{\rm kt~ yr})$.

 Depending on the flux parameters, and on the detector setup,  $\sim 10-26$ events are expected from the \df, indicating a low-to-moderate statistics, similar to that of the observed burst from SN1987A (a total of 20 events observed at Kamiokande and IMB, see \cite{Hirata:1987aa, PhysRevLett.58.1494}). 
  The energy distribution of the signal events shows the features already discussed for the $\barnue$ flux, Sec. \ref{sec:flux}. For JUNO, the Low signal is below the background throughout the energy spectrum; 
  the number of \df\ events is higher than the background in the energy range $\sim$ 12-22 MeV for the Fiducial case and $\sim$ 12-28 MeV for the High signal case. In the realistic energy window, the atmospheric NC background dominates over the atmospheric $\barnue$ CC (see Appendix \ref{C}).  
 The situation is more promising for SLS, where the signal exceeds the background in all cases and for the entire spectrum within the energy window. This is due to the slightly lower background than JUNO, and the much higher signal efficiency, as discussed before. The signal-to-background ratio is $S/B\sim 3$ ($S/B\sim 4$) for the Fiducial (High) case. At lower energies, the atmospheric NC background dominates and then it becomes comparable to the atmospheric $\bar{\nu}_{e}$ CC at about $\sim$ 19-21 MeV, and then the latter starts to dominate at higher energies.  The NC background is much smaller in SLS than in JUNO (see Appendix \ref{C}).  
 
 For  \skg, the background dominates over the signal in all cases and at all energies, by at least a factor of $\sim 1.5$.  Of the $\sim 28$ background events, 16 are due to the NC processes (see Appendix \ref{C}). The signal-to-background ratio is the lowest at intermediate energy, $E_{e^+}\simeq 16-24$ MeV, where the NC atmospheric background and the invisible muon background contribute comparably. Therefore, in our calculations, we chose an energy window of 12-26 MeV as shown in Table \ref{table:3}. For  $E_{e^+} \lta 16$ MeV, the NC atmospheric background becomes dramatically strong. This fact has led us to different, and less promising conclusions compared to earlier phenomenology literature where the detectability of the \df\ in \skg\ was estimated without including NC processes.

\subsection{Detectability prospects}
\label{Detectabilityprospects}
Let us now address the question of the significance of a possible \df\ signal:
how likely is it that the diffuse neutrino flux predicted here will produce a statistically significant excess in a detector?

To answer this question, we employ a hypothesis testing method \cite{Demortier:2014aa},  
 which involves comparing the data with (at least) two different hypotheses. These are $H_0$, the null or background-only hypothesis, 
 and $H_1$, the signal+background  hypothesis, where in this case the signal is due to the DSNB as predicted in Sec. \ref{sec:flux}. For a given detector,  the statistical variable is the number of events in the energy window. We denote $n$  and $n_{obs}$ the ``true'' (from Eq. (\ref{eq:nevents})) and the observed numbers of events respectively. 

Conventionally, the criterion  to claim the evidence of $H_1$ (and therefore evidence of the \df\ according to our model),
is that the probability ($p$-value) that $n_{obs}$ is  realized in the $H_0$ hypothesis be $p<3\times10^{-3}$. This is equivalent to requiring an excess of at least 3$\sigma$ for a Gaussian distribution\footnote{Here  the Poisson statistics is used, however, because it is fully general and applies rigorously to the entire range of number of events of interest here.}. Let $N_{3\sigma}$ be the minimum value of $n_{obs}$ that satisfies this criterion.
 We define the probability of evidence, $P_{ev}$, as the probability that  $n_{obs}> N_{3\sigma}$ is realized in the $H_1$ hypothesis. In intuitive terms, $P_{ev}$ represents the probability that, due to statistical fluctuations in the number of signal+background events, 
 a sufficiently large excess above background is observed in the detector; thus implying evidence of the \df. We note that $P_{ev}$ is larger for larger separation between the $H_1$ and $H_0$ hypotheses, i.e., for larger signal\footnote{As an example,  let us consider a hypothetical situation where signal and background contribute equally, with 11 signal events and 11 background events expected. Then, $n=11$ for  $H_0$ (background only) and $n=22$ for $H_1$ (background + signal).  In this case, $N_{3\sigma}\simeq 21$, and one gets $P_{ev}\equiv P^{eq}_{ev} \simeq 58\%$.  This parameter, $P^{eq}_{ev}$, can be a useful reference to interpret the results of this section; we will find  $P_{ev}< P^{eq}_{ev}$ for the (discouraging) cases where the signal is lower than the background.}.
 
 \begin{table}[htbp]

\begin{center}
\begin{tabular}{ |p{2.3cm}|p{2.3cm}|p{1.1cm}|p{1.1cm}|p{1.1cm}|p{3.04cm}|p{1.2cm}|  }
 \hline
Detectors& Observed energy range (MeV)&NSFCs&BHFCs&Total \df\ &Background (N$_{3\sigma}$)&P$_{ev}(\%)$\\
 \hline
 
 \textbf{Liquid Scintillator}&&&&&&\\

 & &\vspace{2mm} \{4.66\}&  \vspace{2mm}\{1.32\} &\vspace{2mm}\{5.98\}& & \vspace{2mm}\{24.2\}\vspace{2mm}\\
 \multirow{3}{1cm}{JUNO}  &\vspace{2mm}{11-30} &\vspace{2mm} 7.14&  \vspace{2mm}3.04&\vspace{2mm}10.18&\vspace{2mm}{8.02 (17)} & \vspace{2mm}64\\
 
  &  &\vspace{2mm} [7.43]&  \vspace{2mm}[7.53] &\vspace{2mm}[14.96]& & \vspace{2mm}[91.5]\vspace{2mm}\\
  
 \cline{2-7}
  & &\vspace{2mm} \{8.39\}&  \vspace{2mm}\{2.37\}&\vspace{2mm}\{10.76\}&& \vspace{2mm}\{77.6\}\\
\vspace{2mm} \multirow{2}{1cm}{SLS}  &\vspace{2mm}{11-30}  &\vspace{2mm} 12.85&  \vspace{2mm}5.47&\vspace{2mm}18.32&\vspace{2mm}{5.95 (14)} & \vspace{2mm} 98.7\\  

 & &\vspace{2mm} [13.37]&  \vspace{2mm}[13.55]&\vspace{2mm}[26.92]&& \vspace{2mm}[99.7]\\
 
 \hline

 \vspace{0.3mm} \textbf{Water Cherenkov}&&&&&&\\
 \vspace{1mm}
 & &\vspace{2mm}\{4.9\}&  \vspace{2mm}\{1.4\}&\vspace{2mm}\{6.3\}&& \vspace{2mm}\{6.6\}\\

\multirow{2}{1cm}{SuperK-Gd}  &\vspace{1mm}{12-26}  &\vspace{1mm} 7.5&  \vspace{1mm}3.24&\vspace{1mm}10.74&\vspace{1mm}{28.3 (44)} & \vspace{1mm}23 \\

 & &\vspace{2mm}[7.78]&  \vspace{2mm}[8.01]&\vspace{2mm}[15.8]&& \vspace{2mm}[52.3]\\

 \hline
\end{tabular}
\end{center}
\caption{The expected (``true'') number of events for the \df\ (for $\bar{p}=0.68$) and the background, in the observed energy window for \skg\ and a liquid scintillator detector,  with an exposure ${\mathcal M}=225~{\rm kt~ yr}$ and ${\mathcal M}=200~{\rm kt~ yr}$, respectively. The number of signal events  correspond to the Fiducial case (the Low and High signal case are in braces and in square brackets respectively).  $N_{3\sigma}$ is given in round brackets for the background-only hypothesis. The last column gives the probability that a high significance excess due to the \df\ is observed in the detector (see text for details).  }
\label{table:3}
\end{table}

 Table \ref{table:3} gives $n$, $N_{3\sigma}$ and $P_{ev}$ for the three different detector configurations in Sec. \ref{subsec:detectors}, $\bar p=0.68$, and  the Low, Fiducial and High flux cases of Table \ref{Table:2}.  
We find that the results are overall promising, with $P_{ev} > 50\%$ in most cases. 
 For JUNO,  $P_{ev}$ $\sim$ 64\% ($\sim$ 92\%) for the Fiducial (High) signal case.   
 As expected from Sec. \ref{subsec:eventrates}, SLS is even more promising, with $P_{ev} > 77\%$ in all cases.  If we consider the $p$-value commonly required to claim discovery, $p \geq 3 \times 10^{-7}$ (i.e., a  excess of $5\sigma$ or larger), we find that the probability of achieving  it for SLS is $P_{disc}\sim$ 70\% and $P_{disc} \sim$ 98\% for the Fiducial and High cases, respectively.
For SuperK-Gd, the  potential of obtaining a high significance signal is more modest: $P_{ev}\sim 23\%$ 
%
%
and  $P_{ev}\sim 52\%$ for the Fiducial and High signal cases.  Although moderately encouraging, these results may serve as a motivation to further improve the background rejection, especially in the NC channel.

 The results in Table \ref{table:3} give a partial answer to the question of how $P_{ev}$ depends on the uncertainties on the \df\ model.  The range of $P_{ev}$ becomes broader if one also varies $\bar p$ in the interval $\bar p=0 - 0.68$ (see Appendix \ref{B}): specifically, we find the ranges: $P_{ev}=(14.5-91.5)\%$ (for JUNO), $P_{ev}=(53.4-99.7)\%$ (for SLS), and $P_{ev}=(4.1-52.3)\%$ (for \skg). 
To further characterize uncertainties, we also computed $P_{ev}$ for other flux models taken (with minor approximations) from different literatures, specifically for the DA08+M08 model in   \cite{0004-637X-804-1-75} and the SN 1987A \& BH model in   \cite{YUKSEL2015413}. These were  chosen as representative of extreme values of the flux, so they give correspondingly extreme values of $P_{ev}$. For \skg, we find $P_{ev}\sim 6\%$ for the DA08+M08 model and $\sim 52\%$ for the SN 1987A \& BH model. For JUNO, the corresponding values are  $P_{ev}\sim 20\%$  and $\sim 93\%$.   

In addition to a single detector performance, it is interesting to consider the joint potential of two detectors,  like \skg\ and JUNO,  to establish evidence of the \df\ as predicted in our model.
Let us begin by denoting as $p^{\scalebox{.5}{\rm{JUNO}}}_i(n^{\scalebox{.5}{\rm{JUNO}}}_{obs} ~|~H_i)$ the Poisson probability that JUNO registers $n^{\scalebox{.5}{\rm{JUNO}}}_{obs}$ events in the hypothesis $H_i$.  A similar definition holds for $p^{\scalebox{.5}{\rm{SK}}}_i(n^{\scalebox{.5}{\rm{SK}}}_{obs}~|~H_i)$.        

In terms of these single-detector probabilities, the combined probability of observing the number of events $n^{\scalebox{.5}{\rm{SK}}}_{obs}$ and $n^{\scalebox{.5}{\rm{JUNO}}}_{obs}$ in the hypothesis $H_i$ is
\begin{equation}
L_{i}(n^{\scalebox{.5}{\rm{SK}}}_{obs},n^{\scalebox{.5}{\rm{JUNO}}}_{obs})=p^{\scalebox{.5}{\rm{SK}}}_i(n^{\scalebox{.5}{\rm{SK}}}_{obs}~|~H_i) p^{\scalebox{.5}{\rm{JUNO}}}_i(n^{\scalebox{.5}{\rm{JUNO}}}_{obs}~|~H_i)~.
\end{equation}
In order of compute the probability that the total signal (combined of the two detectors) is significant over the background, we consider two conditions: 
\beq
&& L_1(n^{\scalebox{.5}{\rm{SK}}}_{obs},n^{\scalebox{.5}{\rm{JUNO}}}_{obs})\geq10^{-4}
\label{condition1} \\
&& \frac{L_0(n^{\scalebox{.5}{\rm{SK}}}_{obs},n^{\scalebox{.5}{\rm{JUNO}}}_{obs})}{L_1(n^{\scalebox{.5}{\rm{SK}}}_{obs},n^{\scalebox{.5}{\rm{JUNO}}}_{obs})} \leq10^{-3}  ~.
\label{condition2}
\eeq
 Thus the joint probability of evidence is then defined as
 \be
 P^{(2)}_{ev}=\sum L_1(n^{\scalebox{.5}{\rm{SK}}}_{obs},n^{\scalebox{.5}{\rm{JUNO}}}_{obs})~,
 \label{pev2}
 \ee
where the summation is over all the pairs $(n^{\scalebox{.5}{\rm{SK}}}_{obs},n^{\scalebox{.5}{\rm{JUNO}}}_{obs})$  that satisfy the conditions (\ref{condition1}) and (\ref{condition2}). Eq. (\ref{condition1}), is a high likelihood condition: it means that the joint probability of observing a certain pair $(n^{\scalebox{.5}{\rm{SK}}}_{obs},n^{\scalebox{.5}{\rm{JUNO}}}_{obs})$ in the $H_1$ hypothesis is sufficiently large to make the hypothesis $H_1$ credible. We checked that the probability that a pair falls in the region identified by Eq. (\ref{condition1}) is about 98\%. 
The second condition, Eq. (\ref{condition2}), is on the likelihood ratio: it requires that  the same pair of numbers of events is much more likely to be realized in the $H_1$ hypothesis than in $H_0$, so that $H_1$ would be a favored interpretation of this observation. 
Combining the two conditions physically means choosing those pairs $(n^{\scalebox{.5}{\rm{SK}}}_{obs},n^{\scalebox{.5}{\rm{JUNO}}}_{obs})$ that have a reasonably high probability to be realized in $H_1$ and at the same time a fairly low probability in the $H_0$ hypothesis.

\begin{figure}[htbp]

\begin{minipage}[c]{7.8cm}
\includegraphics[width=1 \textwidth]{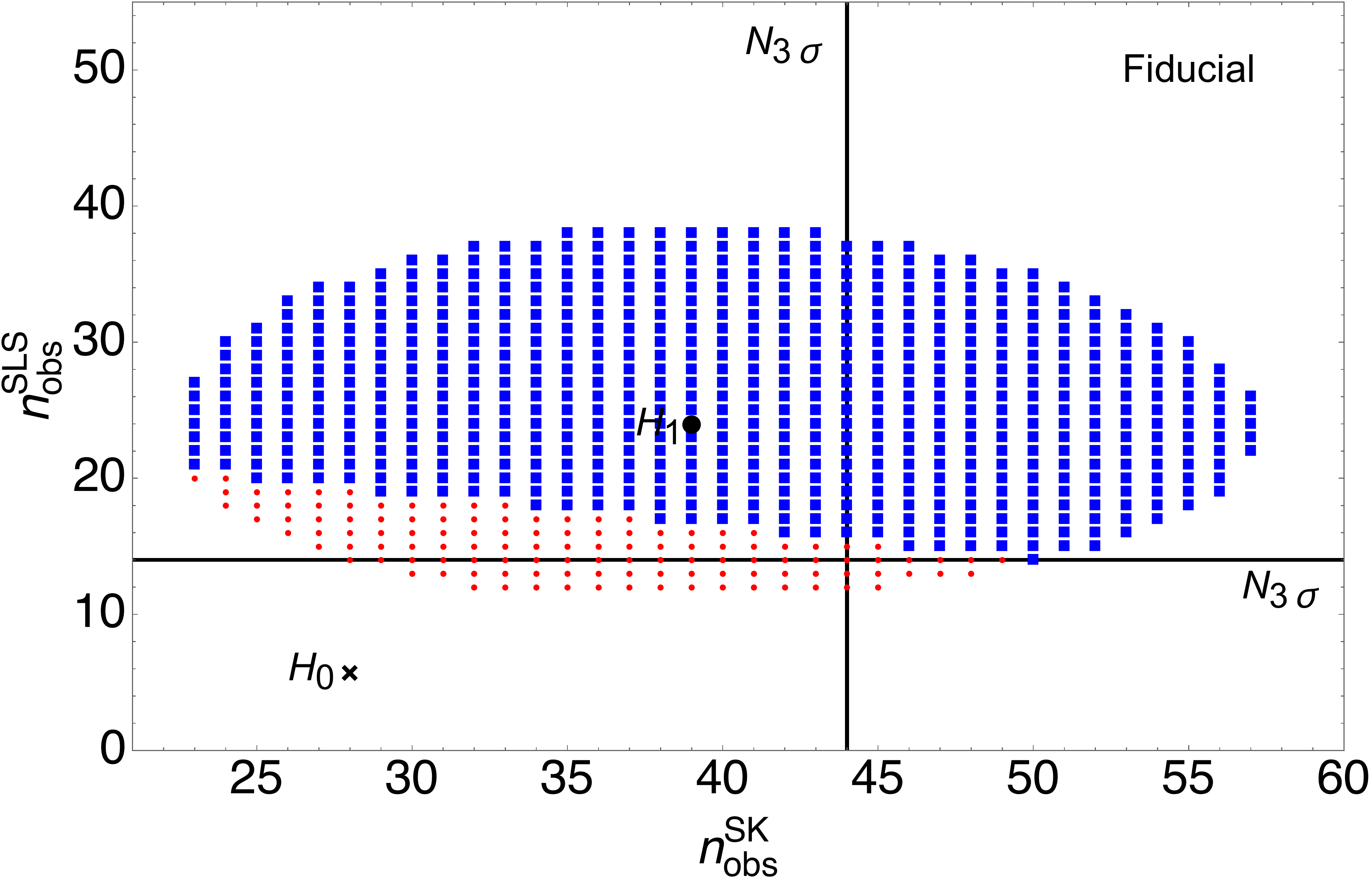}
\end{minipage}
\begin{minipage}[c]{7.8cm}
\includegraphics[width=1 \textwidth]{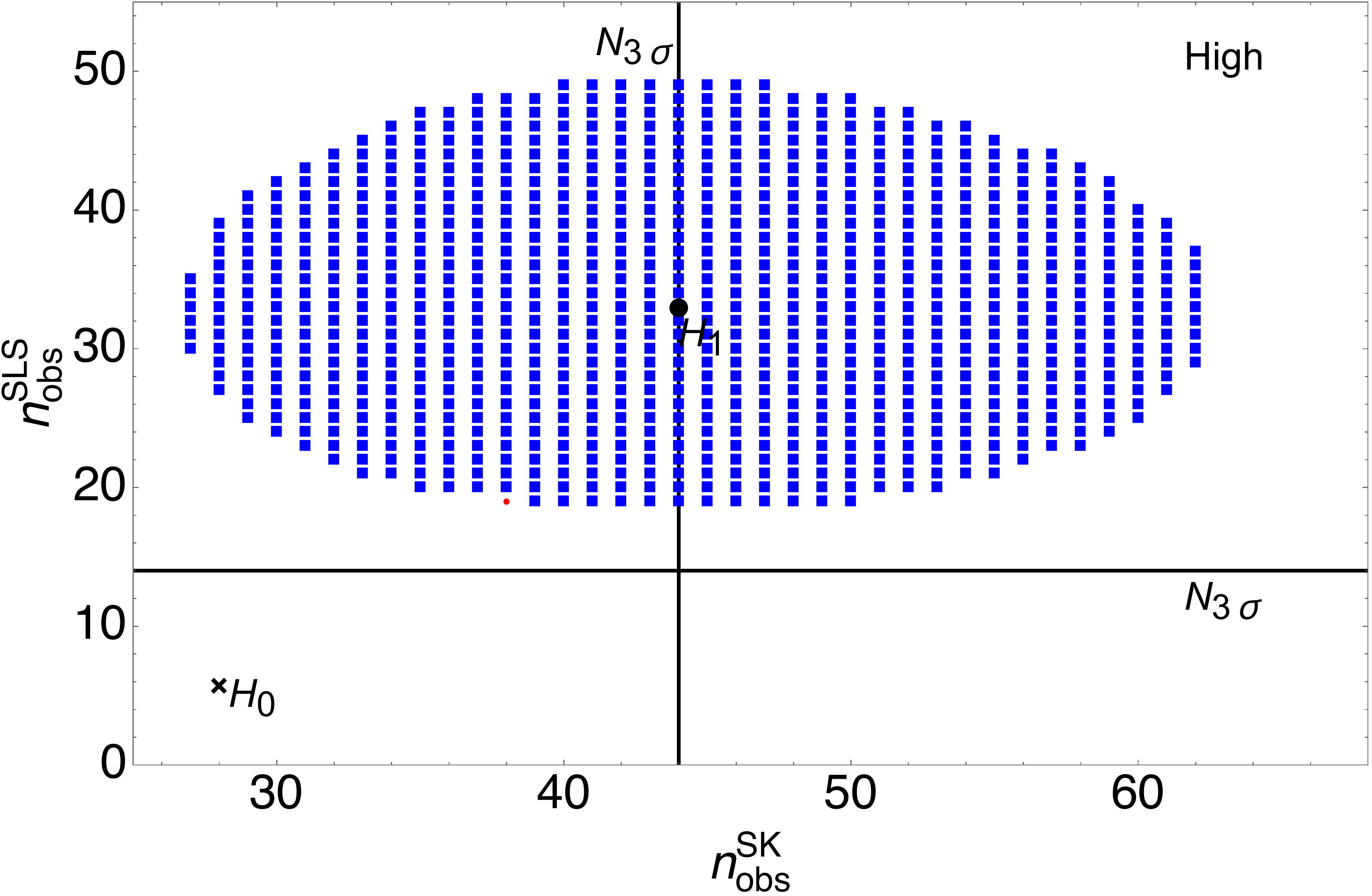}
\end{minipage}

\begin{minipage}[c]{7.8cm}
\includegraphics[width=1 \textwidth]{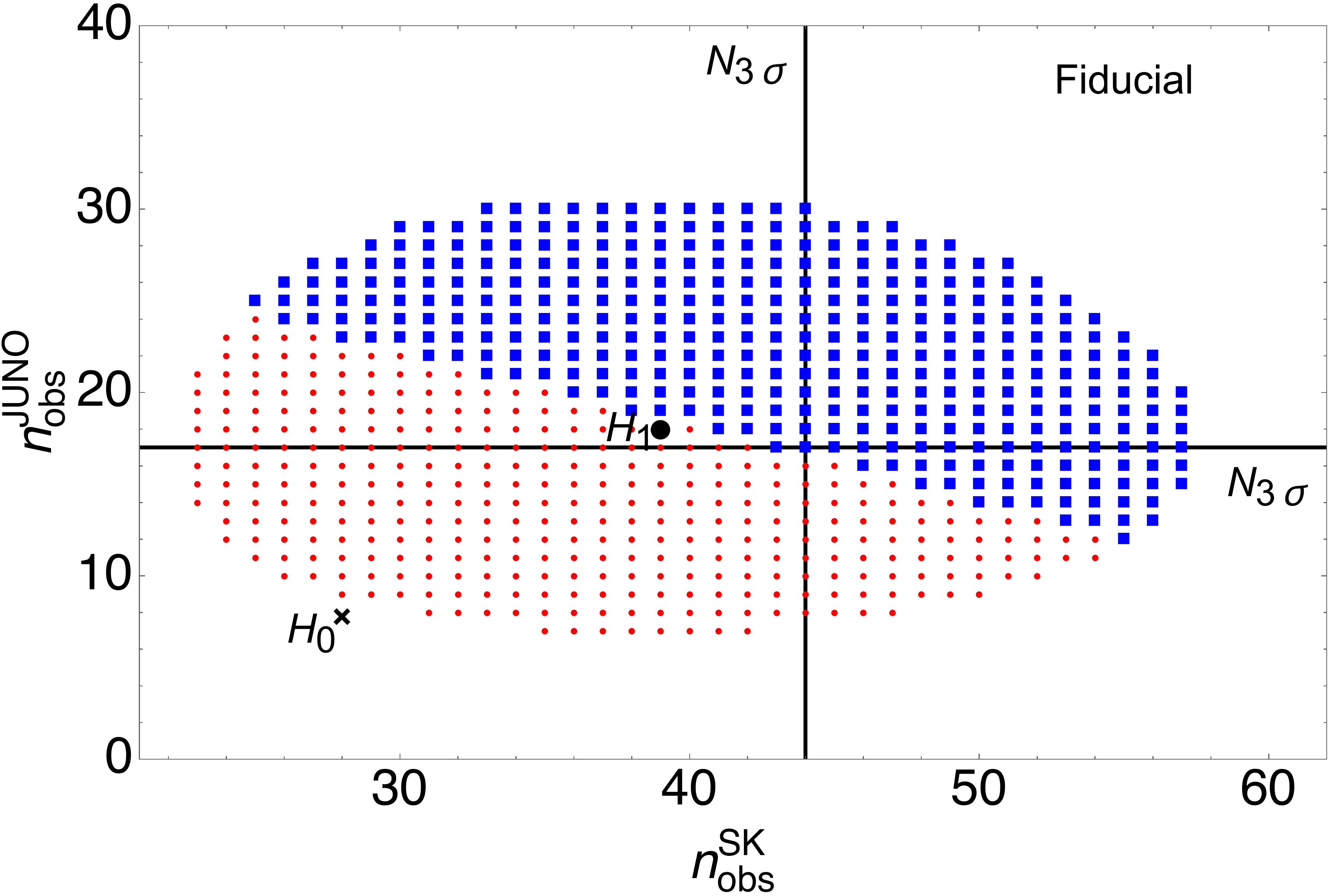}
\end{minipage}
\begin{minipage}[c]{7.8cm}
\includegraphics[width=1 \textwidth]{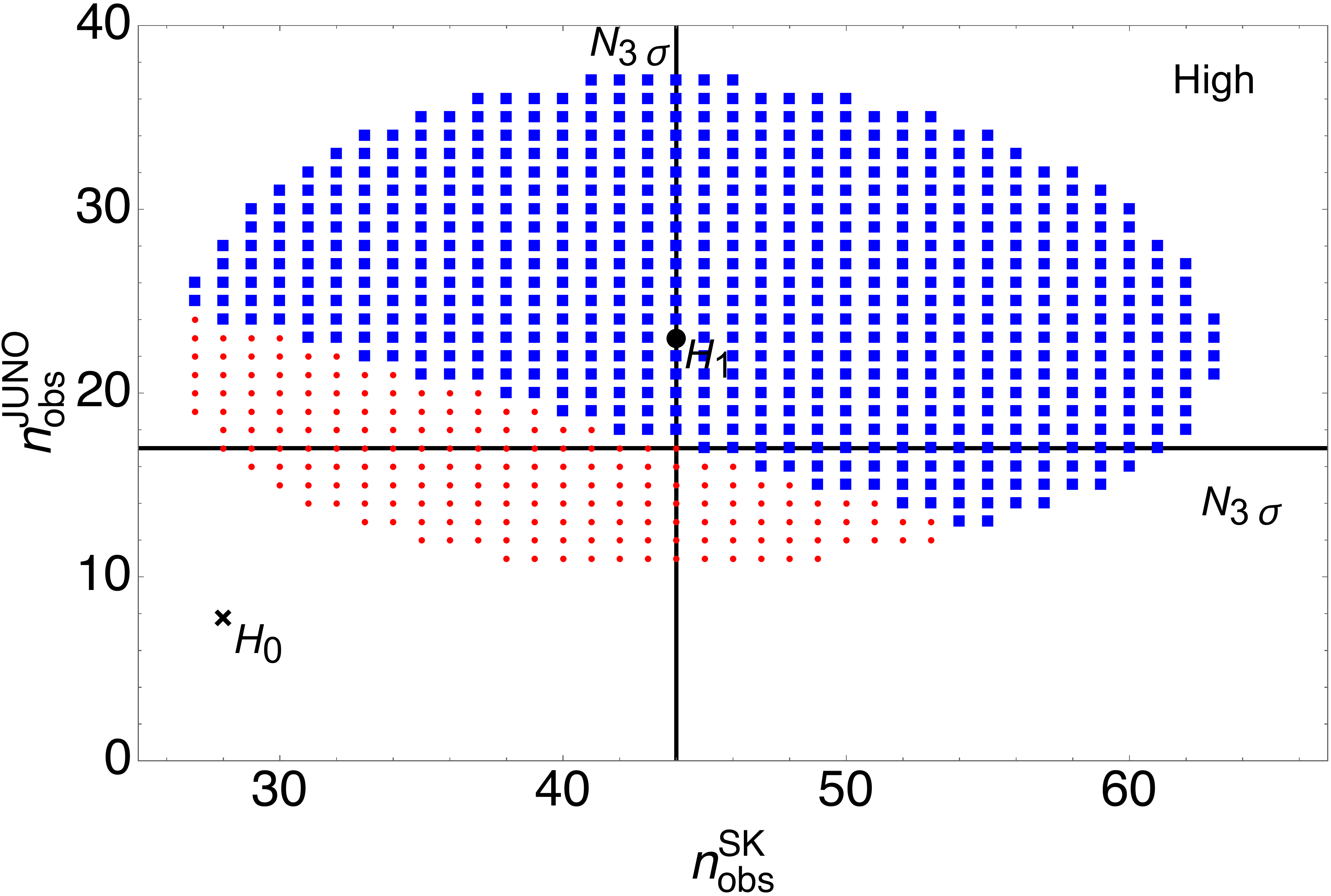}
\end{minipage}
\caption{\label{figure6*} The ``true'', predicted numbers of events for \skg\ and SLS or JUNO for background only ($H_0$ hypothesis, black cross) and signal$+$background ($H_1$ hypothesis, black dot), for the cases in Table \ref{table:3}.  The horizontal and vertical lines mark $N_{3\sigma}$, which is the number of events corresponding to a 3$\sigma$ excess in $H_0$ hypothesis (see text).   Also shown is the region where only the high likelihood condition (Eq. (\ref{condition1}))  is satisfied (dots, red) and where both the high likelihood and the likelihood ratio conditions  (Eqs. (\ref{condition1}) and  (\ref{condition2})) are fulfilled (squares, blue).
}
\end{figure} 
In Fig. \ref{figure6*} we show the region in the space of $(n^{\scalebox{.5}{\rm{SK}}}_{obs},n^{\scalebox{.5}{\rm{JUNO}}}_{obs})$ where only the condition (\ref{condition1})  is satisfied (dots, red) and where both conditions  (\ref{condition1}) and  (\ref{condition2}) are fulfilled (squares, blue). 
 The figure also shows the  ``true'', predicted values of the numbers of events for the individual detectors in the $H_0$ and $H_1$ hypotheses, and the values of $N_{3\sigma}$ for each detector.  The figure gives results for both the combination of \skg\ and  JUNO and of \skg\ and SLS, for the Fiducial and High signal cases (Table \ref{table:3}).

Confirming the results in Table \ref{table:3}, Fig. \ref{figure6*} makes it clear that the configuration with SLS is the most promising: for the combination of \skg\ and SLS and the Fiducial signal flux, we find $P^{(2)}_{ev}\simeq$ 92\%.   For the High signal case,  the entire region of high likelihood has high likelihood ratio; the corresponding probability of evidence is $P^{(2)}_{ev}\simeq$ 98\%.

For the combination of \skg\ and JUNO we find more conservative results, mostly due to the lower signal efficiency of JUNO (relative to SLS):  $P^{(2)}_{ev}\simeq$ 44\% ($P^{(2)}_{ev}\simeq$ 84\%), for the Fiducial (High) case.

\section{Discussion and conclusions}
\label{sec:discussion}

We have presented an updated study of the Diffuse  Supernova Neutrino Background (\df) and its short-to-medium term detection prospects.    The \df\ is modeled using the results of state-of-the art numerical simulations of the Garching group for both direct black hole-forming collapses and neutron star-forming collapses.  Three different scenarios for the dependence  of the collapse outcome (black hole or neutron star) on the mass of the progenitor star are presented, corresponding to a fraction of black-hole-forming collapses between $\sim 10\%$ and $\sim 30\%$.  The progenitor dependence of the  \n\ flux is included as well. The detection potential of ${\mathcal O}(10)$ kt  liquid scintillator and water \ck\ (with Gadolinium) detectors is assessed, using the most detailed estimates for the background processes, including neutral-current scattering of atmospheric neutrinos. 
 
Let us summarize our main results. 

\begin{itemize}

\item The diffuse $\barnue$ flux in a detector, for $E \geq 11$ MeV of neutrino energy, should be $\phi_{\barnue} \simeq (1.4 - 3.7)~{\rm cm^{-2}s^{-1}}$, where the interval was obtained by varying the progenitor dependence of black hole formation, the normalization of the core collapse rate, and the flavor conversion probability.
 This is a factor of $\sim 3 - 9$ below the current \sk\ bound \cite{PhysRevD.85.052007}.
 Therefore, an improvement by about one order of magnitude in experimental sensitivity is required to guarantee detection. Depending on the progenitor dependence of the collapse outcome (fig. \ref{figure1}), and on the neutrino energy, the contribution of black-hole-forming collapses to the total  $\barnue$ flux ranges between minor ($\sim$20\% or less for $E \sim 11-30$  MeV) and dominant (up to $\sim 70$\% for $E \gta 30$ MeV).

\item  We calculated the event rate expected in JUNO for the detector performance described in \cite{0954-3899-43-3-030401}.
If our flux prediction is accurate (and if the background is modeled with negligible uncertainty), there is a probability up to  $P_{ev} \sim 90\%$  that an excess of events in JUNO will be established above a $\sim 3\sigma$ significance in a decade of operation.  
The signal due to the \df\ should be comparable to or larger than the background at least for $11 < E_{vis} < 17$ MeV. 

For Slow Liquid Scintillator, as proposed in \cite{Wei2017255}, the signal should exceed the background in nearly the entire energy window. We obtain $P_{ev} \simeq 98\%$ for our Fiducial set of parameters and the same exposure as JUNO. For the golden standard of discovery, a $5\sigma$ excess, we find a $\sim 70\%$ probability for the same parameters.

\item  
  for \skg,  the event rate is  dominated by the background at all energies, with  the neutral current  scattering of atmospheric \ns\ being the dominant background at  $E \lta 16$ MeV. 
We find that  the potential to observe a statistically significant excess is severely limited,
with $P_{ev} \lta 52\%$ in all cases for ten years of operation.  

\item We estimated the probability that the flux  predicted here will produce a statistically significant (likelihood ratio $\sim 10^{-3}$ or smaller) signal in \jn\ and \skg\  when the two experiments are considered jointly. We find that this probability could be  $\sim 45-85\%$ depending on the parameters.
  If Slow Liquid Scintillator is used in combination with \skg, the probability exceeds $\sim 90\%$ for typical \df\ parameters.

\end{itemize}

We stress that our results are effected by a number of uncertainties, the largest one being on the normalization of the rate of core collapses. This uncertainty is not fully understood, and scenarios with a rate higher than the range considered here are not excluded. 
Another uncertainty is on the current understanding of the backgrounds at water and liquid scintillator detectors. It is possible that conclusions will change as these backgrounds become better-known.

Although with the limitations described above, we can draw two broad conclusions.  The first is that the contribution of failed \sne\ to the \df\ can be substantial. 
This conclusion, already found in earlier literature, remains true after  including the most updated information on failed \sne, their progenitor stars, and their \n\ emission.
The potential to use \ns\ to probe the birth of black holes is especially interesting. It could contribute to the new era of multimessenger studies that has been  pioneered by the recent detection by LIGO-VIRGO \cite{PhysRevLett.116.061102} of gravitational waves from a merger of stellar-mass black holes, which could be the remnants of failed \sne.

The second main message is that the potential of the short-medium term \n\ experimental program to observe the \df\ is  strong, although less so than previously anticipated in early studies, where backgrounds were not fully accounted for. Advanced techniques of background discrimination -- especially those that allow a high efficiency for the signal --  will be critical for success, and should be mainly targeted to improving the discrimination of atmospheric neutral current backgrounds. In this respect, the use of  LAB as a Slow Liquid Scintillator, possibly with wavelength shifters to enhance its performance, seems especially promising. For water with Gadolinium, it is in principle possible to further reduce the backgrounds by devising more stringent topological cuts (to be used in addition to the Cherenkov angle selection cut, see Sec. \ref{subsec:water}). Efforts are planned on this within the \skg\ collaboration \cite{Vagins}. We find that in the ideal case of complete subtraction of neutral current backgrounds, $P_{ev}$ could exceed $\simeq 90\%$.

To conclude, there is a realistic possibility that the \df\ will be discovered within a decade or so, thus delivering a unique and direct picture of the landscape of collapsing stars. There is hope that the potential to extract information on the rate of collapses, neutrino transport inside the star, and black hole formation, will motivate a strong and sustained experimental effort to increase the sensitivity even further, and ultimately transition from discovery to precision studies in the longer term.

\acknowledgments

We thank H. T. Janka and A. Summa for providing the results of the numerical simulations of the Garching group, and for useful discussions. We are also grateful to H. Kunxian, M. Wurm and J. Hidaka for informative comments. We acknowledge support from the Department of Energy award DE- SC0015406.


\appendix

\section{Appendix: input neutrino spectra}\label{A1}

In this Appendix, details of the \n\ fluxes and spectra used in this work are given.
We use numerical results by the Garching group \cite{phdthesis,Mirizzi:2015eza}, which were obtained for solar mass progenitors (taken from \cite{Woosley:2002aa}) in conditions of spherical symmetry, using the Lattimer and Swesty equation of state \cite{LATTIMER1991331}.  The numerical runs, the progenitor masses and the outcomes of the collapse (\nsf\ or \dbh)  are listed in Table \ref{Table:1}.

%
%
The Garching group simulations contain state-of-the-art treatment of the neutrino transport, including processes such as neutrino-pair conversion between different flavors, energy transfer in neutrino-nucleon interactions, and nucleon-correlations in the dense medium \cite{phdthesis}. The multi-dimensional effects of convection are taken into account in an effective way, via a mixing-length treatment \cite{phdthesis,Mirizzi:2015eza}\footnote{Initial results from multidimensional simulations, where convection is treated more realistically, show only minor differences in the \n\ spectra and luminosities compared to the quantities used in this work. See, e.g., \cite{phdthesis,Mirizzi:2015eza}.}.
%
For \nsf,  the simulation time extends up to $\sim 10$s, and therefore the cooling of the proto-neutron star is included. For \dbh, the simulations reproduce the expected duration of the \n\ burst of ${\mathcal O}(1)$s.

For each \n\ species, $w$, the output files gives the  time dependent luminosity $L_w$,
the average energy $\langle E \rangle_w$, and the second energy moment $\langle E^2 \rangle_w$. 

\begin{figure}[htbp]
 \begin{tabular}{cc}
 \includegraphics[width=0.50 \textwidth]{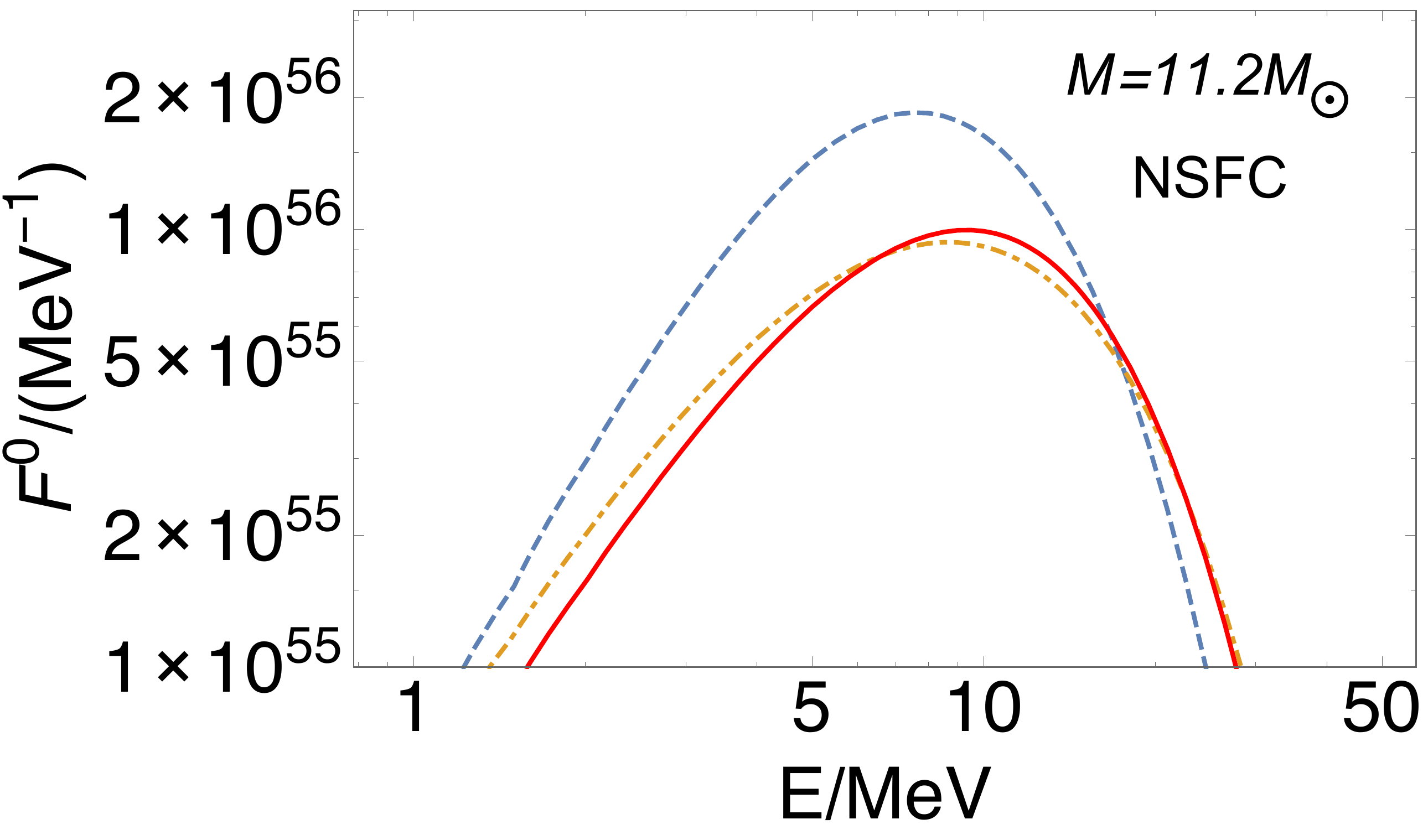}&\\
\includegraphics[width=0.50 \textwidth]{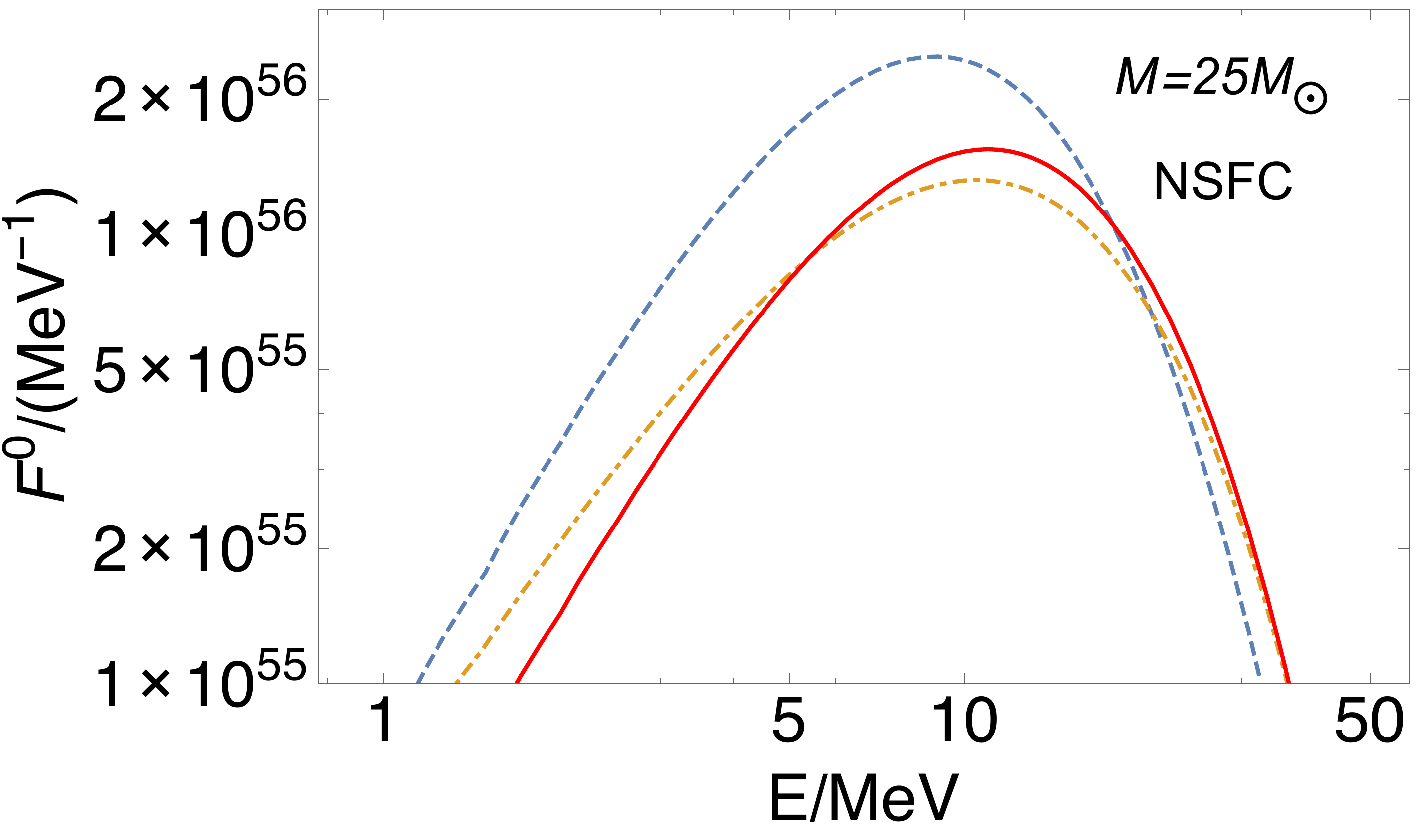}&
 \includegraphics[width=0.50 \textwidth]{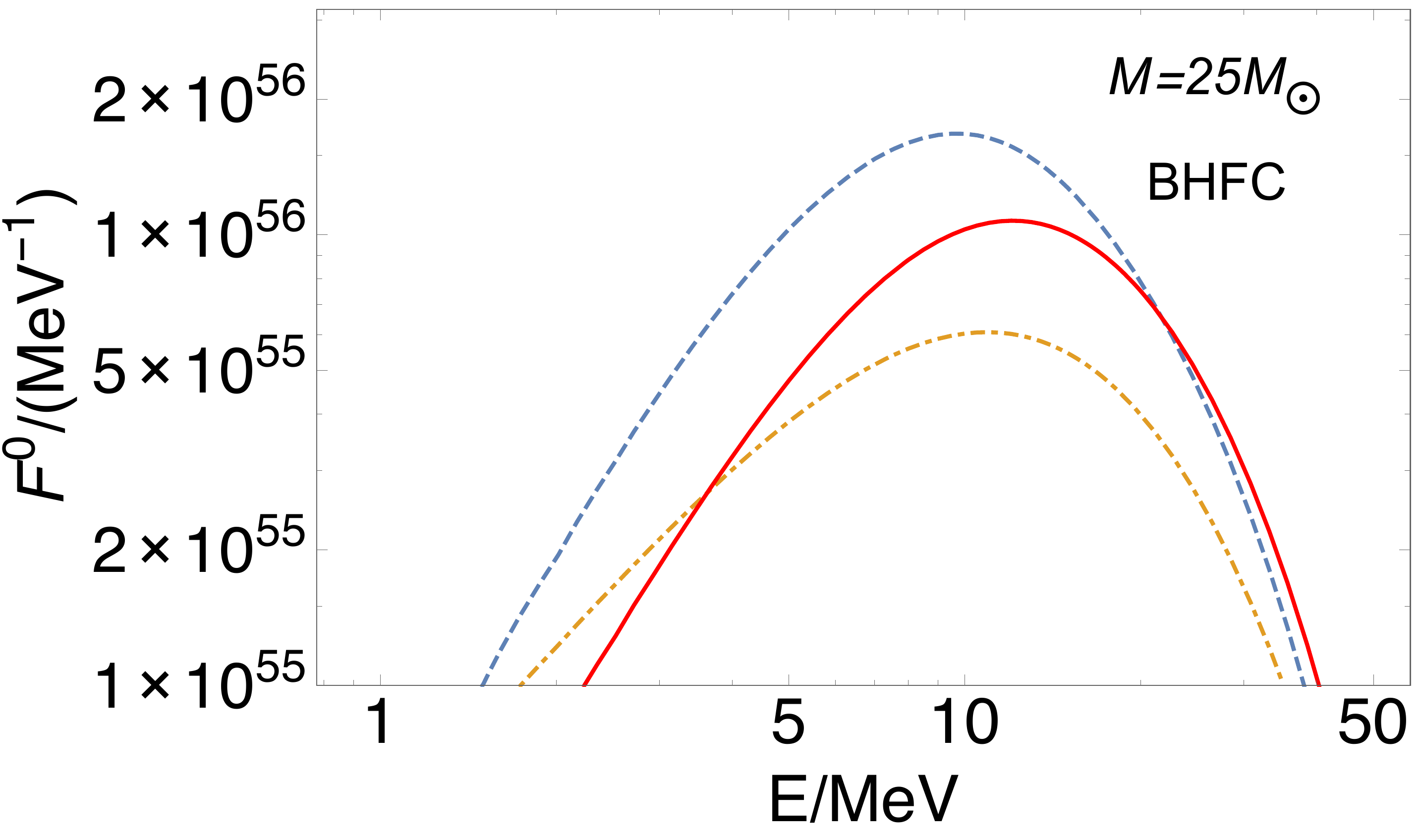}\\
 \includegraphics[width=0.50 \textwidth]{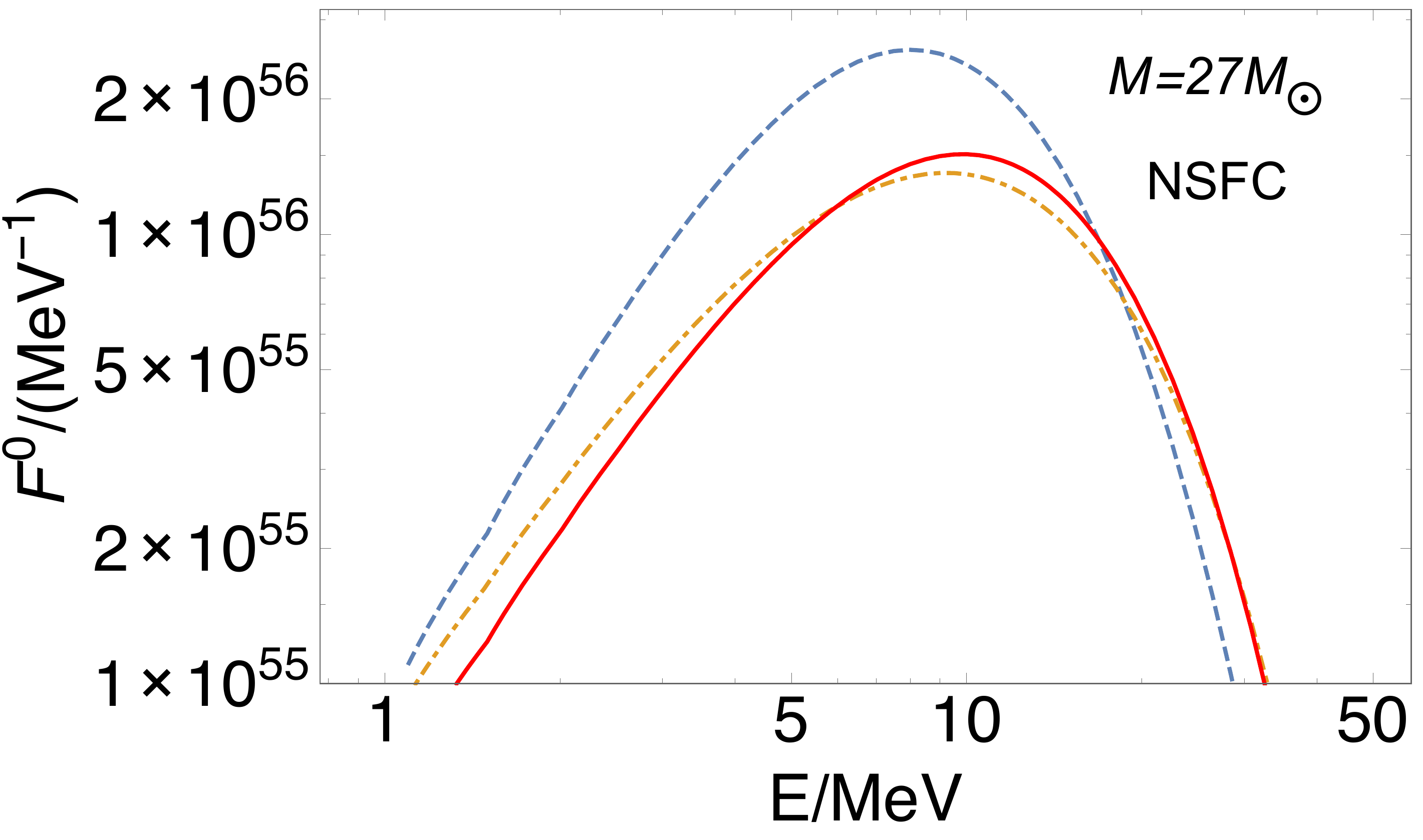}&\\
 &\includegraphics[width=0.50 \textwidth]{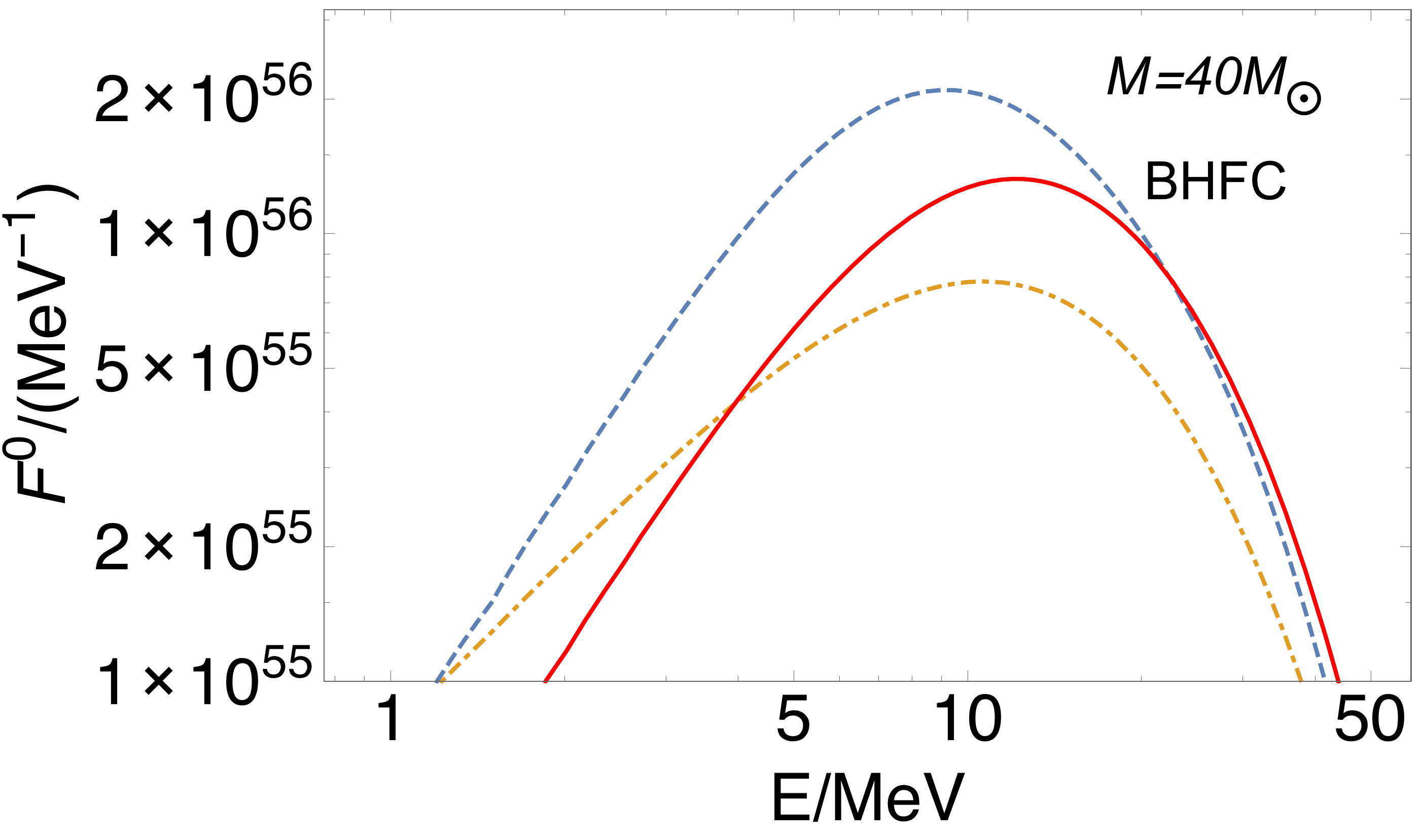}
 \end{tabular}
 \caption{\label{figure2} The time-integrated spectra at production (before oscillations), for different \n\ species and different progenitor masses, $M$. The panes are ordered vertically with increasing $M$ (legends). Left column: successful explosion (\nsf); right column: Black-hole forming collapses (\dbh). The blue dashed,  red solid and yellow dot-dashed lines correspond to the $\nu_e$, $\bar{\nu}_e$ and $\nu_x$ spectra, respectively. }
\end{figure}

At each instant of time, the energy spectrum is 
modeled as suggested in \cite{0004-637X-590-2-971}:     
\begin{equation}\label{eq: 2.4}
f^0_{w}(t)=\frac{L_w}{\langle E\rangle^2_w} \frac{(1+\alpha_w)^{(1+\alpha_w)}}{ \Gamma(1+\alpha_w)}.\Big(\frac{E}{\langle E \rangle_w }
\Big)^{\alpha_w} e^{-(\alpha_w+1)E/ \langle E \rangle_w}~,
\end{equation}
 where
\begin{equation}
\alpha_w=\frac{\langle E^2\rangle_w-2\langle E\rangle^2_w}{\langle E\rangle^2_w -\langle E^2 \rangle_w}~.
\end{equation}
From this equation, using the time-dependent parameters above, the time-integrated flavor spectra were obtained. They are described in Table \ref{Table:1} and Fig. \ref{figure2} in terms of the total energy emitted per flavor, $\mathcal{L}_w$, and first two energy moments, $\langle \epsilon \rangle_w$ and $\langle \epsilon^2 \rangle_w $. 

From Table \ref{Table:1} we note that for \dbh\  $\langle \epsilon \rangle_w$ is $\sim 20\%$ higher than for \nsf, and the emission of energy is stronger for $\nue$ and $\barnue$ than for $\nux$, due to the high rate of electron and positron capture on nuclei in the hot matter accreting on the collapsed core \cite{Sumiyoshi:2006aa}. 
 For the two \dbh\ simulations used here, with $M=25\msun$ and $M=40\msun$, the flavor spectra are nearly identical, but a $\sim 30\%$ larger energy output is realized overall for the more massive progenitor.  

For the \nsf\ simulations, a 
 non-monotonic behavior of the parameters is observed with the increase in progenitor mass:  this is not surprising,  as the properties of the explosion are not directly related to $M$, but rather depend strongly on the stellar structure, mass loss rate, etc. \cite{0004-637X-762-2-126,0004-637X-757-1-69,0004-637X-818-2-124}. In particular, 
 it was found that  the pre-explosion neutrino emission depends on the compactness parameter \cite{0004-637X-762-2-126}, which is a non-monotonic function  of the progenitor mass.


\begin{table}[htbp]
 \begin{center}
\begin{tabular}{ p{0.8cm}c c c c c c c c c c c c c}
\hline\hline
 \multicolumn{1}{c}{} & 
    \multicolumn{7}{c}{} 
 \\
Run (Type)&Mass/$M_{\odot}$ &$\mathcal{L}_{\nu_e}$&$\mathcal{L}_{\bar{\nu}_e}$ &$\mathcal{L}_{\nu_x}$&&$\langle \epsilon \rangle_{\nu_e}$&$\langle \epsilon \rangle_{\bar{\nu}_e }$&$\langle \epsilon\rangle_{\nu_x}$&&$\langle \epsilon^2 \rangle_{\nu_e}$&$\langle \epsilon^2 \rangle_{\bar{\nu}_e }$&$\langle \epsilon^2\rangle_{\nu_x}$\\
 \cline{3-5}
 \cline{7-9}
 \cline{11-13}\\
&& \multicolumn{3}{c}{$[10^{52}$ ergs]} &&
  \multicolumn{3}{c}{[MeV]}&&\multicolumn{3}{c}{[MeV$^2$]} \\
  \hline\hline\\
s11.2c (NSFC) &11.2&3.56 &3.09 & 3.02&&10.43&12.89& 12.93&&137.52&213.18&220.86\\
 s25.0c (NSFC) &25&7.18& 6.78&6.02 && 12.67&15.5&15.41&&209.19&310.2&315.35 \\
s25.0c (BHFC) &25&7.08 & 6.51&3.7 &&15.32&18.2&17.62&&318.92&437.57&427.22 \\
s27 (NSFC)&27& 5.87&5.43 &5.1&&11.3&13.89&13.85&&164.68&249.97&255.38 \\
s40.0c (BHFC)&40&9.38 & 8.6&4.8 &&15.72& 18.72&17.63&&343.65&470.76&440.71\\
\hline
\end{tabular}
\caption{Summary of the numerical results from the Garching group \cite{phdthesis,Mirizzi:2015eza} used in this work. For each \n\ species, the table gives the total energy emitted and the first two energy moments (i.e, the averages of the energy and of the square of the energy) of the time-integrated spectrum.   All runs use progenitors of solar metallicity from Woosley et al. \cite{Woosley:2002aa}, with the Lattimer and Swesty Equation of State (LS220) \cite{LATTIMER1991331}.   }
\label{Table:1}
\end{center}
\end{table}

\section{Appendix: parameter dependence of the \df}\label{B}

In this Appendix details are given on the variation of the \df\ with the input parameters. 

\begin{figure}[h]
\begin{minipage}[c]{7.4cm}
\includegraphics[width=1 \textwidth]{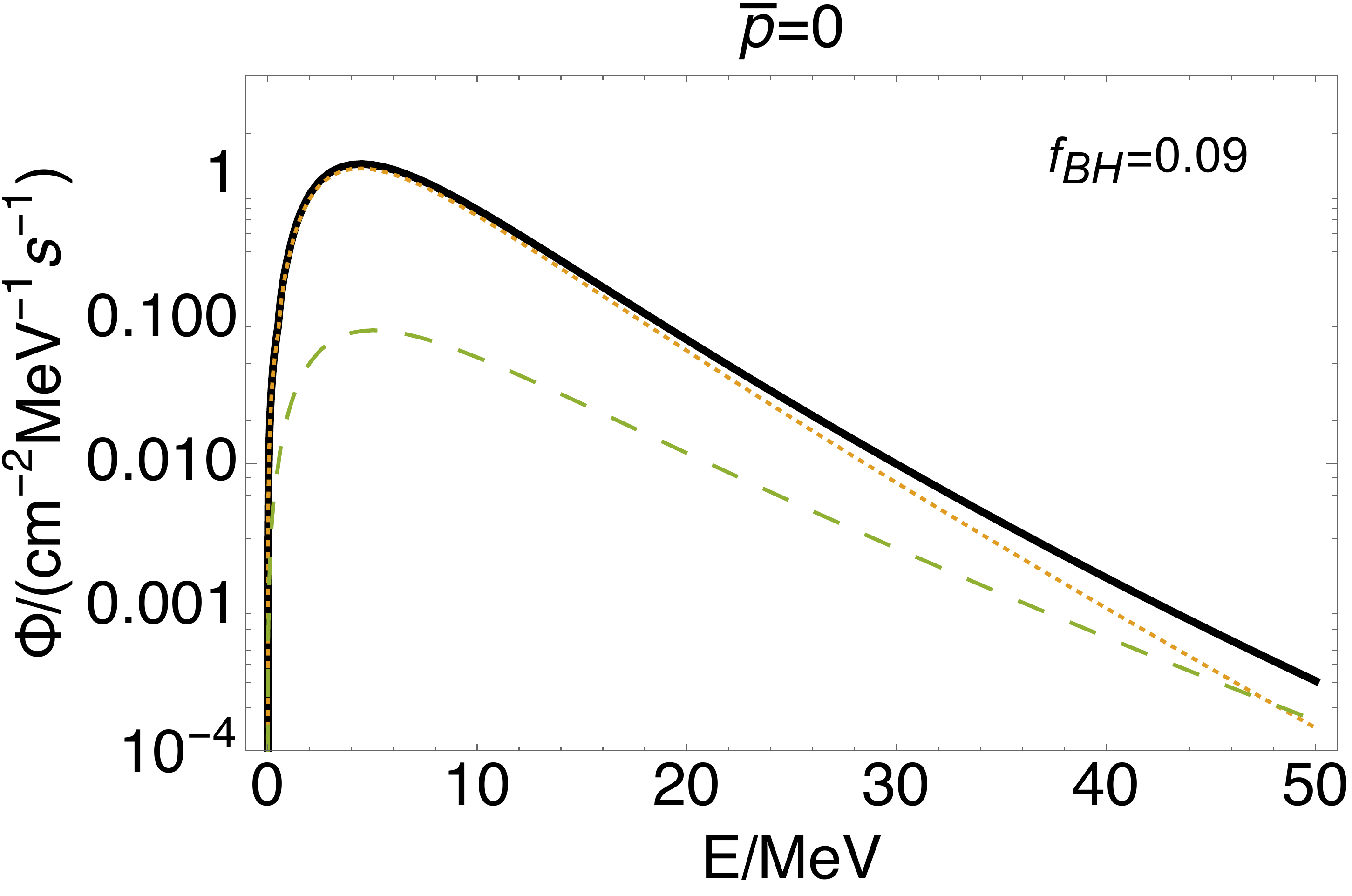}   
\end{minipage}
\begin{minipage}[c]{7.4cm}
\includegraphics[width=1 \textwidth]{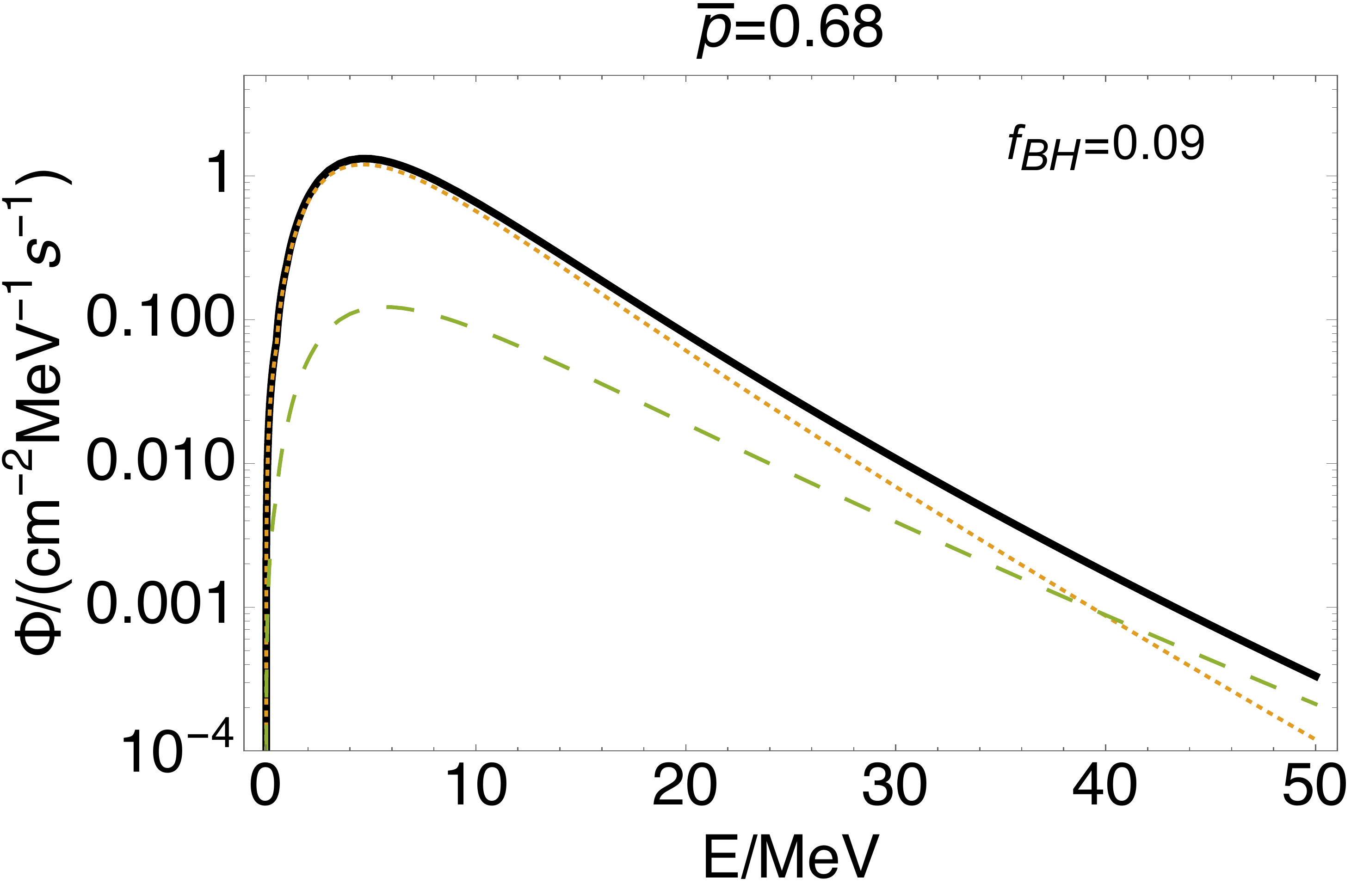}
\end{minipage}

\begin{minipage}[c]{7.4cm}
\includegraphics[width=1 \textwidth]{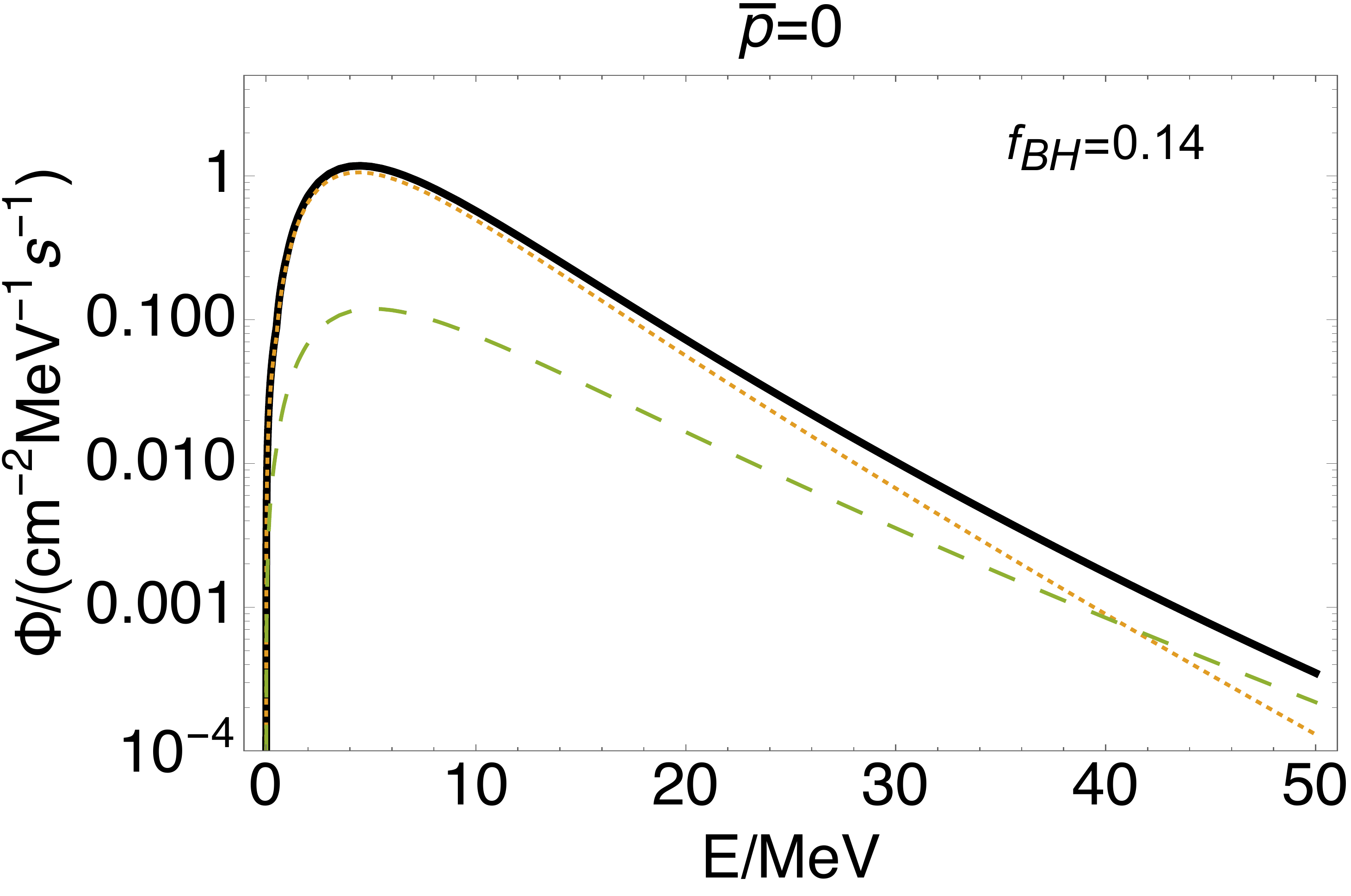}
\end{minipage}
\begin{minipage}[c]{7.4cm}
\includegraphics[width=1 \textwidth]{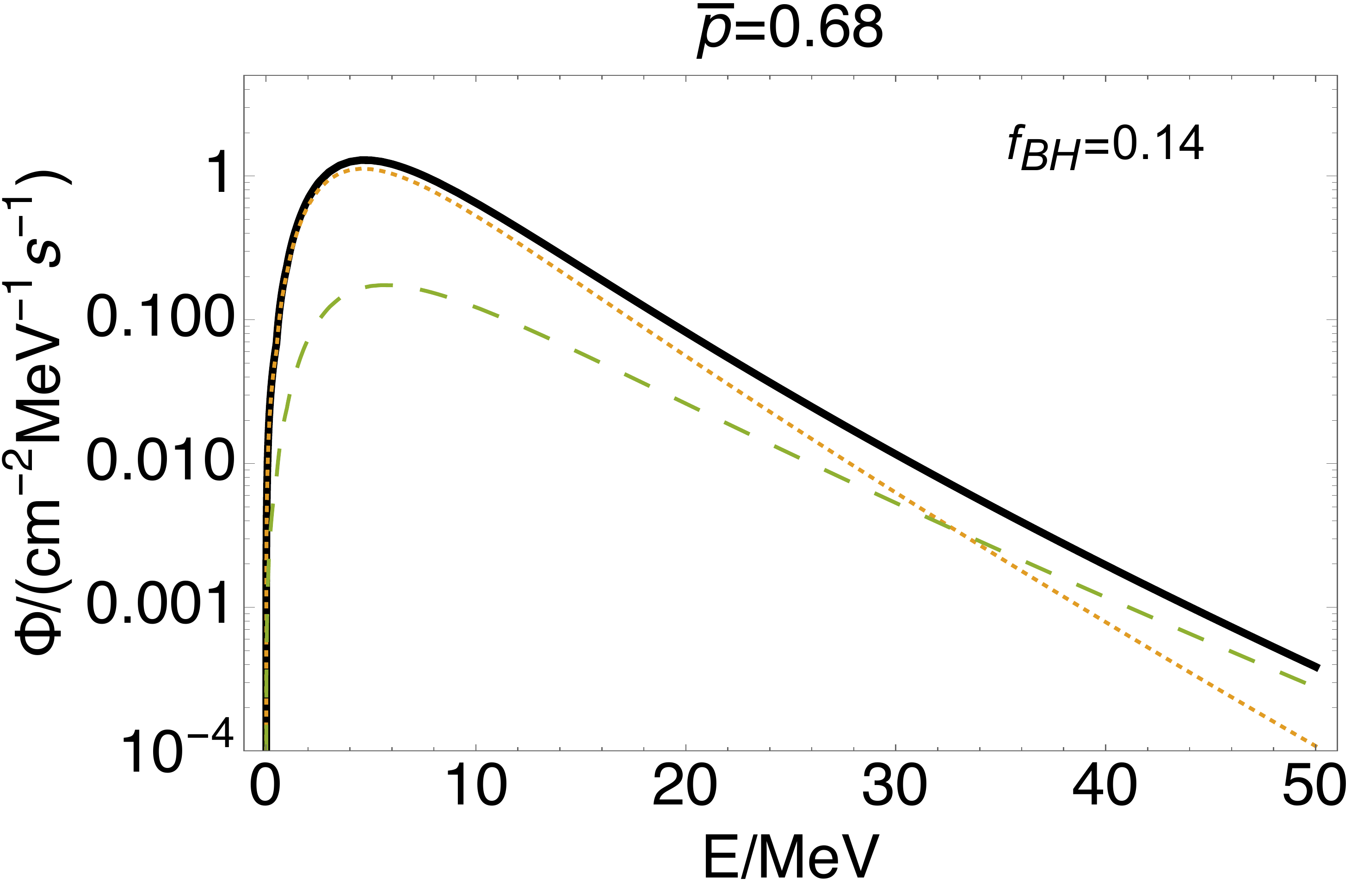}
\end{minipage} 

\begin{minipage}[c]{7.4cm}
\includegraphics[width=1\textwidth]{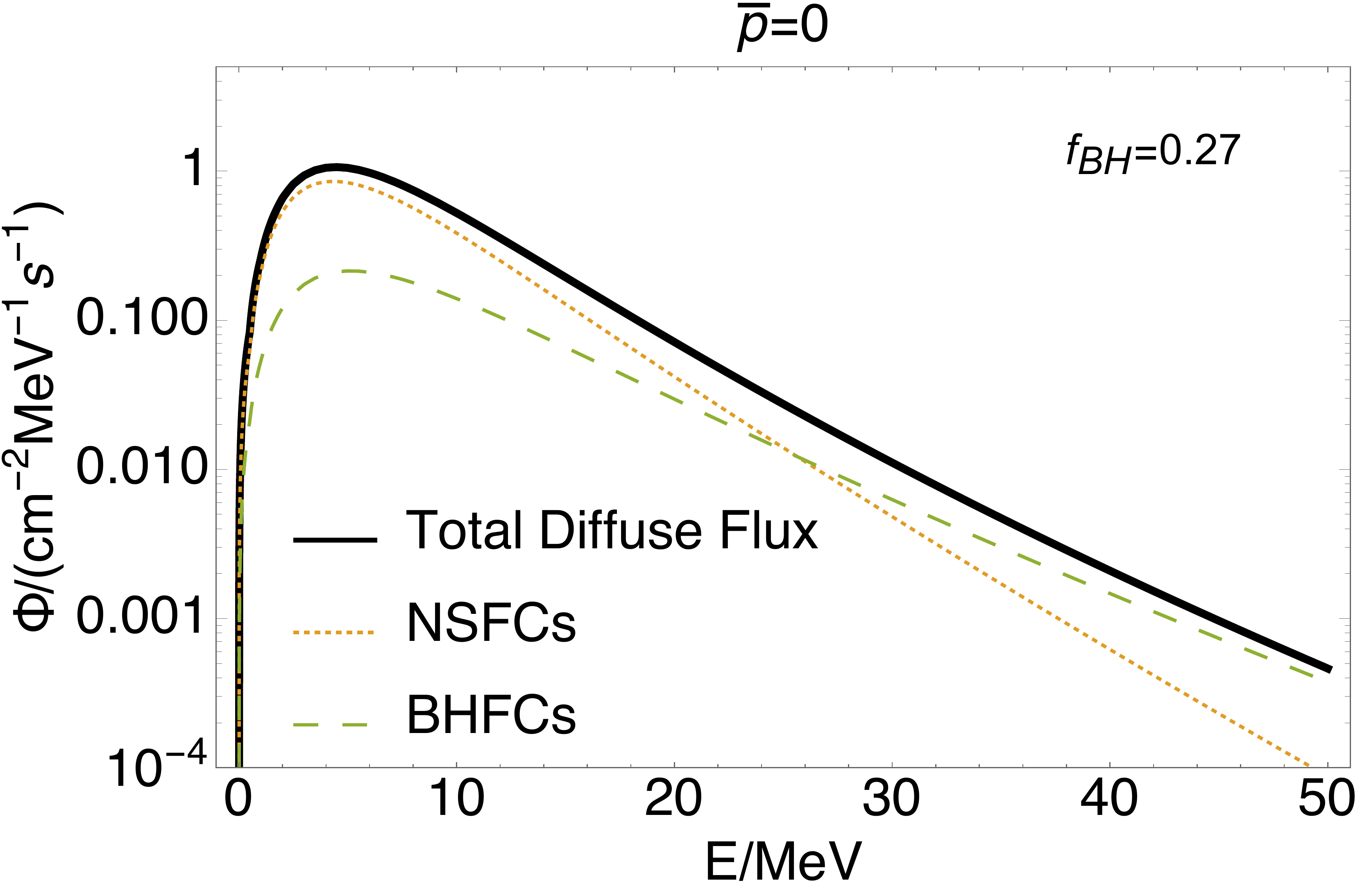}
\end{minipage}
\begin{minipage}[c]{7.4cm}
\includegraphics[width=1 \textwidth]{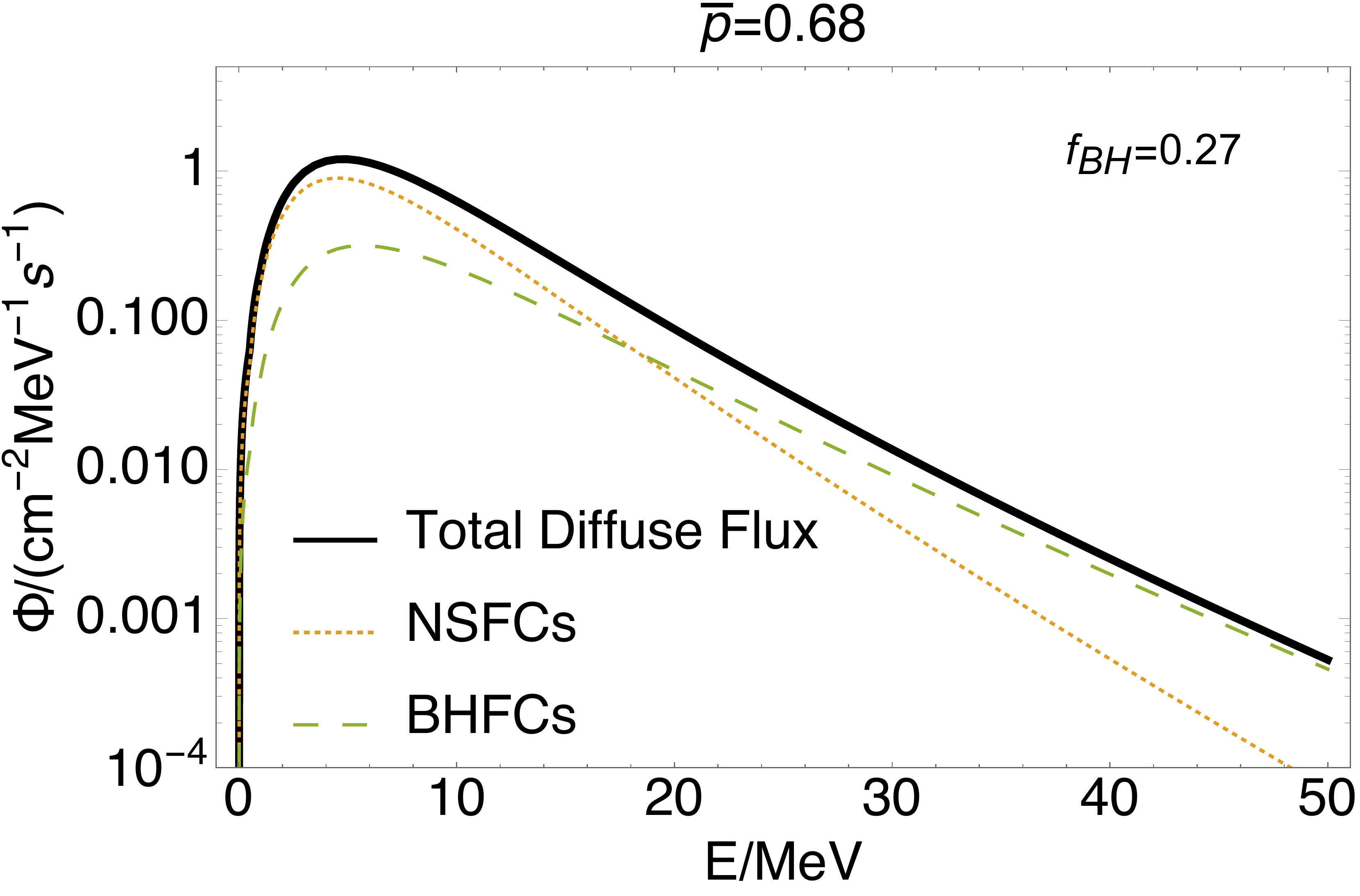}
\end{minipage}

\hspace{\fill}
\caption{\label{figure4} The diffuse fluxes for different scenarios in Fig. \ref{figure1} (labeled by the corresponding fraction of \dbh), assuming a fixed star formation rate, $R_{cc}(0)=$ 1.25 $\times10^{-4}$$\rm{yr^{-1} }$$\rm{Mpc^{-3}}$. Note that these results refer to the redshift bin $z<z_{max}=2$ (see Sec. \ref{sec:flux}). }
\end{figure}

In Fig. \ref{figure4} we show the diffuse $\barnue$ flux, for different survival probabilities $\bar p$,  and different scenarios of dependence of \bh\ formation on the star's progenitor mass, $M$ (see Sec. \ref{sub:progenitors} and Fig. \ref{figure1}).
A fixed core collapse rate is assumed, $R_{CC}(0)=1.25\times10^{-4} \rm{yr^{-1} }\rm{Mpc^{-3}}$ \cite{Lien:2010aa}.  

The figure exhibits a number of expected features of the \df: a peak at $E \sim 5$ MeV, where $\Phi \sim 1~{\rm cm^{-2} s^{-1} MeV^{-1}}$, with an approximately exponential decline at higher energies.  The contribution of \nsf\ is always dominant near the peak energy, while the flux due to \dbh\ becomes increasingly important with increasing energy, due to its hotter spectrum. For the cases with $f_{BH}=0.09, 0.14$, the \dbh\ flux is comparable or larger than the \nsf\ one for $E \gta E_{t}=35- 45$ MeV.
For the case with $f_{BH}=0.27$, the \dbh\ flux dominates above $ E_{t}= 18-26$ MeV. 
This transition energy, $E_t$, falls inside the realistic energy window of detection,
thus suggesting that the effect of failed \sne\ on the \df\ might be detectable.

Overall, the dependence on the oscillation pattern (the $\barnue$ survival probability $\bar p$) is moderate, with variations of the flux at the level of $\sim 20\%$, in the energy interval $E \gta 11$ MeV, when varying $\bar p$. In all cases, $E_t$ decreases by $\sim 6-10$ MeV (indicating stronger \dbh\ dominance in the flux) when $\bar p$ increases from 0 to 0.68. This is expected, considering that for \dbh\ the emission is strongest in the electron flavors (see Table \ref{Table:1}).   

\section{Appendix: backgrounds}\label{C}

\begin{figure}[ht]
 \begin{center}
\includegraphics[width=0.5 \textwidth]{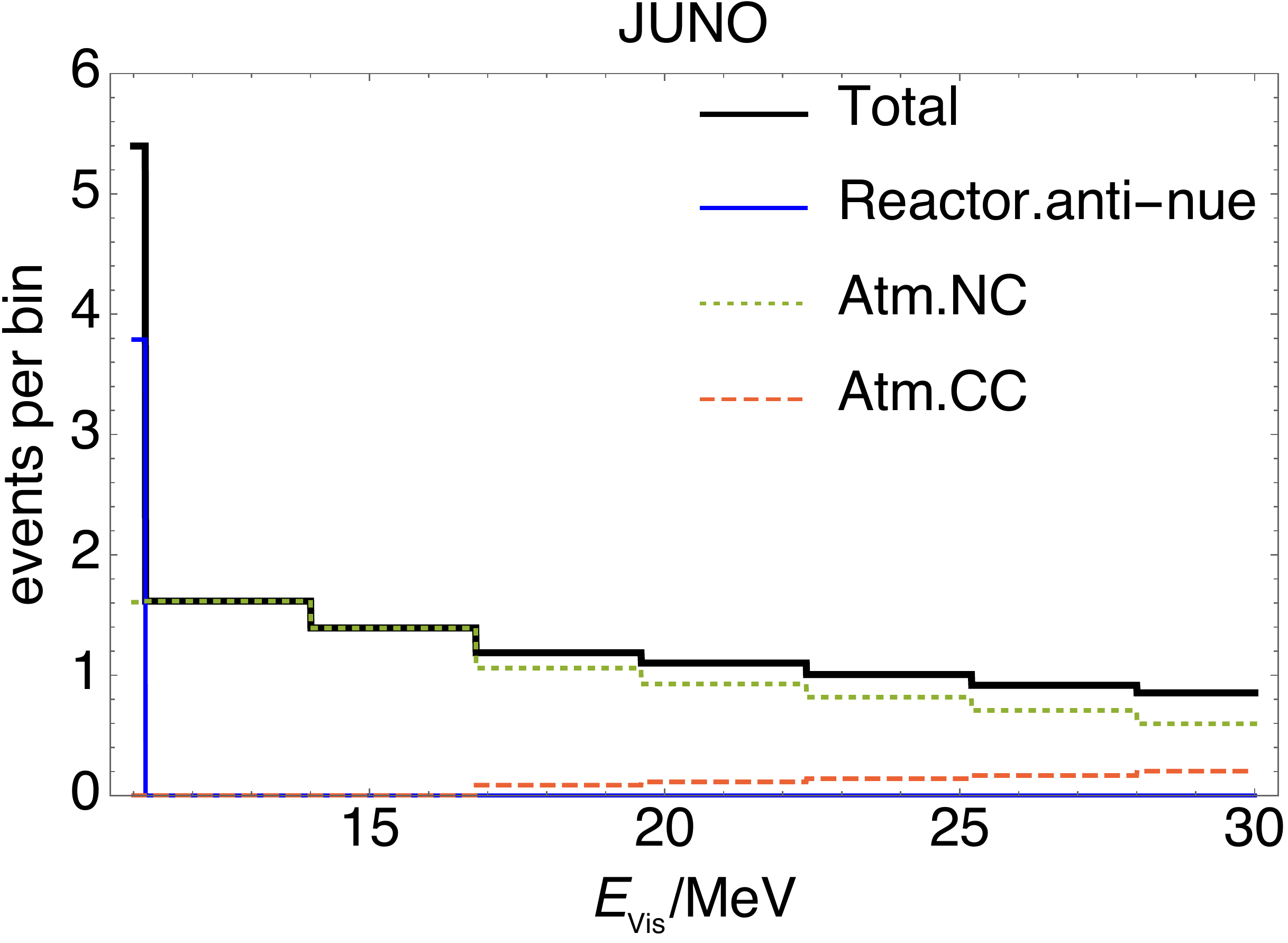}\\
\vskip0.4truecm
\includegraphics[width=0.5 \textwidth]{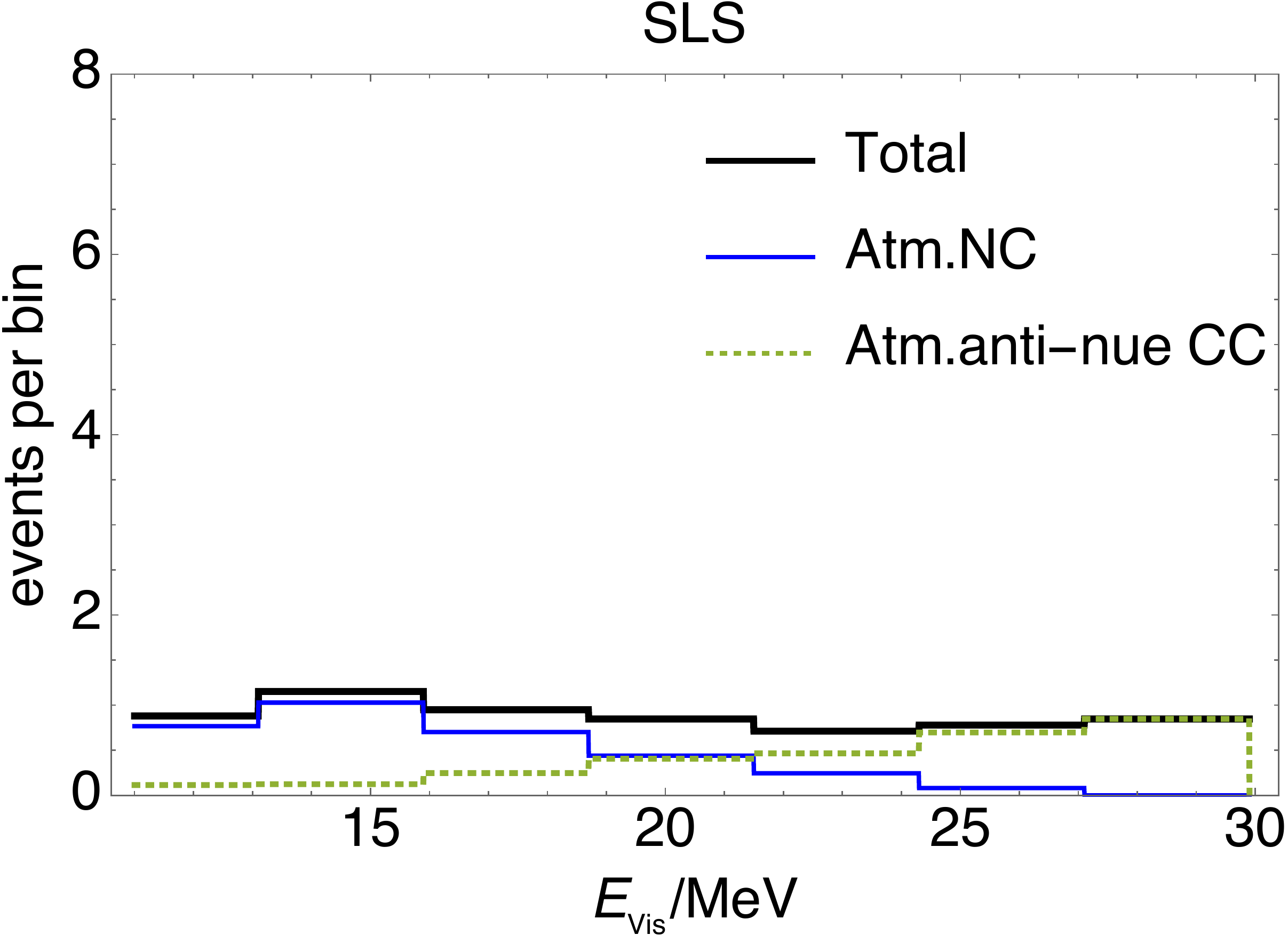}\\
\vskip0.4truecm
\includegraphics[width=0.5 \textwidth]{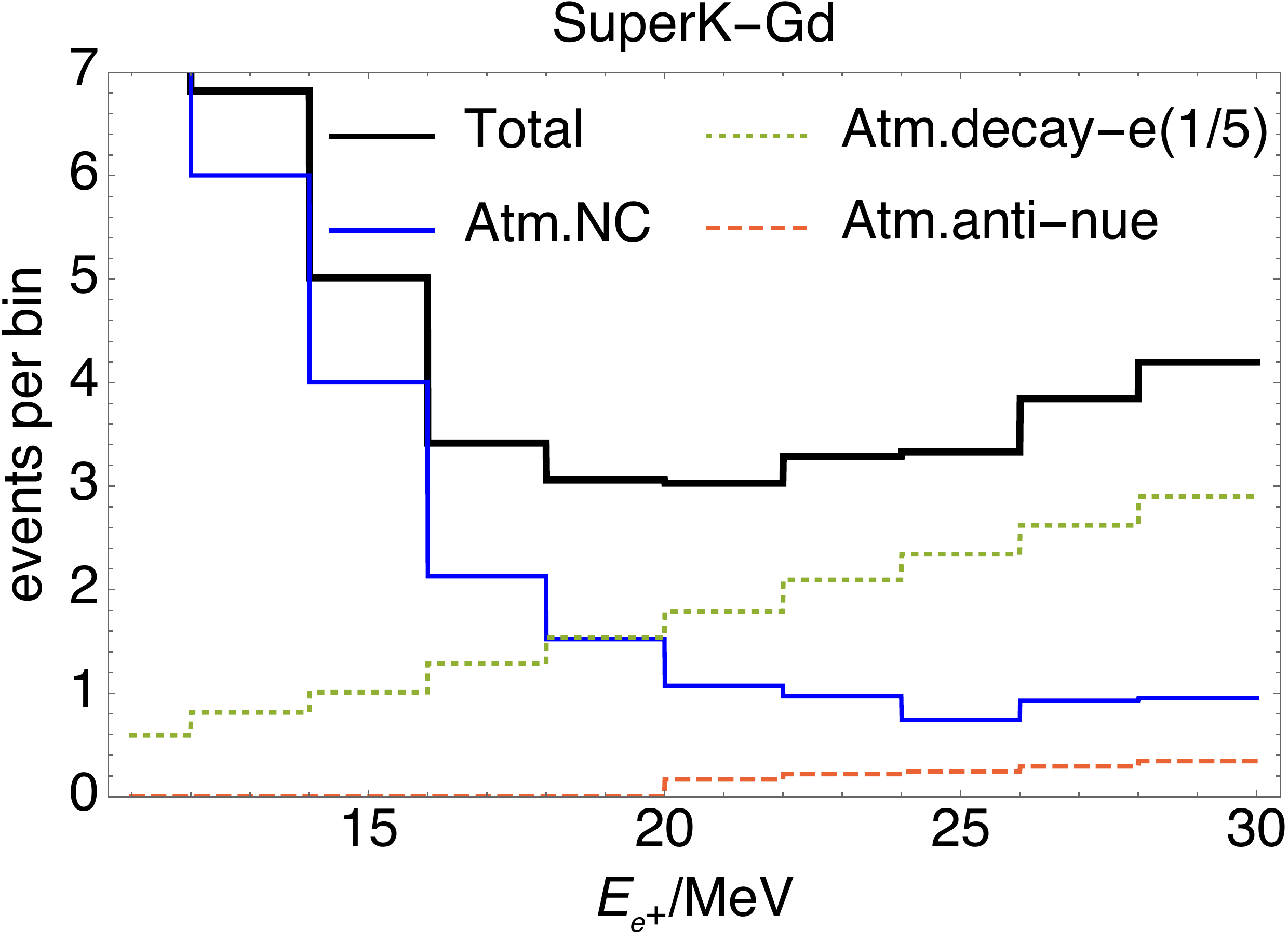}
\end{center}
\hspace{\fill}
\caption{\label{figure9} Different components of the backgrounds in JUNO, SLS and SuperK-Gd. In the bottom pane, the background due to the decay of sub-Cherenkov atmospheric muons (dotted line) is calculated assuming a factor of 5 reduction compared to pure water Cherenkov.}
\end{figure}

In this Appendix we show (Fig. \ref{figure9}) the contribution of different sources and processes to the backgrounds in JUNO \cite{0954-3899-43-3-030401}, SLS \cite{Wei2017255} and SuperK-Gd \cite{phdthesis1}. This content supplements the discussion
 in Sec.  \ref{subsubsec:lab} and Sec. \ref{subsec:water}.

\bibliographystyle{JHEP}
\bibliography{reference}

\end{document}